\documentclass[reprint,eqsecnum,floats,aps,amsmath,amssymb,nofootinbib,prd,onecolumn,showpacs]{revtex4-2}
\usepackage{bm}
\usepackage{graphicx,physics}
\usepackage{amsmath,amssymb,mathtools,mathrsfs}
\usepackage{hyperref}
\usepackage{graphicx}
\usepackage{arydshln}
\usepackage{xcolor}
\usepackage{braket}
\usepackage{tensor}
\usepackage{enumitem,array,textcomp}

\begin{document}

\title{Polar perturbations in Kantowski-Sachs spacetimes and hybrid quantum cosmology}

\author{Guillermo A. Mena Marug\'an}
\email{mena@iem.cfmac.csic.es}
\author{Andr\'es M\'inguez-S\'anchez}
\email{andres.minguez@iem.cfmac.csic.es}
\altaffiliation{Affiliated to the PhD Program, Departamento de F\'isica Te\'orica, Universidad Complutense de Madrid, 28040 Madrid, Spain}
\affiliation{Instituto de Estructura de la Materia, IEM-CSIC, C/ Serrano 121, 28006 Madrid, Spain}

\begin{abstract}
Increasing attention has been recently devoted to the study of Kantowski-Sachs spacetime as a way to explore the interior of a Schwarzschild black hole. In this work, we construct a Hamiltonian formulation for polar perturbations of this spacetime in the presence of a perturbative massless scalar field. Our analysis is based on a truncated action at quadratic order in perturbations. Both background and perturbative degrees of freedom are treated dynamically, forming a combined system that is endowed with the canonical structure obtained from our truncated action. First-order perturbations of the metric and the matter field are described using perturbative gauge invariants, linear perturbative constraints, and their corresponding canonical variables. For the quantum description, we adopt a hybrid approach where the background is quantized with loop quantum gravity techniques and the perturbations with conventional quantum field methods.

\end{abstract}

\maketitle

\section{Introduction \label{sec: I}}

Perturbation theory finds a wide range of applications in general relativity (GR). Small deviations from known solutions allow us to assess how stable and sensitive the gravitational system is to changes. Additionally, it facilitates the study of complicated scenarios where exact solutions are difficult to obtain, providing approximated descriptions within a certain domain of validity. While perturbative approximations can, in principle, extend to any order, in practice only the first few contributions are normally considered. Exact solutions are often impractical, and higher-order terms are typically neglected under the assumption that their impact is minimal. First-order perturbations are expected to capture the essential physics of the system while simplifying its description, which is why most studies focus on them. Overall, perturbation theory is a versatile tool that plays an important role in a variety of gravitational phenomena, such as certain regimes of black hole evolution or their emission of gravitational waves \cite{GW1,GW2,Ep1,Ep2}. 

In the mathematical foundations of perturbation theory in GR, one begins by considering two manifolds, of which one is a well-established reference, the background spacetime, while the other consists of a perturbative variation of this background, which we refer to as the perturbed spacetime. The theory proceeds by comparing the two. However, their relationship is inherently nonunique owing to the absence of a preferred point-to-point identification between them. This ambiguity is known as the gauge problem in perturbation theory \cite{SW}. To attain a consistent description of the perturbations, it is essential to identify quantities that remain invariant regardless of how the two spacetimes are related. These quantities are known as perturbative gauge-invariants and encode the physical information about the perturbations \cite{Bardeen,Mukhanov,Sasaki,RW}. Identifying them may be challenging, but in some respects, the Hamiltonian formulation of GR offers a useful intuitive approach \cite{M,Lan}. 

Integrating GR and quantum mechanics in a common framework remains an open problem in physics. Among current approaches to quantum gravity, loop quantum gravity (LQG) stands out as a promising candidate. LQG addresses this challenge by developing a nonperturbative and background-independent quantization of Einstein’s theory. Rather than relying on the standard metric formulation, it utilizes $SU(2)$ connections and their corresponding canonical momenta, given by (densitized) triads. From these variables, holonomies of the connection along closed paths (loops) and fluxes of the densitized triad through surfaces are constructed. These holonomies and fluxes form the algebra of functions that is quantized \cite{A&L,Thiemann}. A few years after the development of LQG, a subfield known as loop quantum cosmology (LQC) emerged, sharing the foundational principles of LQG but applied to cosmological models. Spacetimes with singularities became a primary focus in LQC, as they represent extreme scenarios where GR breaks down. By leveraging symmetries at the classical level, LQC greatly simplifies the quantum analysis. One of the most extensively studied models in LQC is the Friedman-Lemaître-Robertson-Walker (FLRW) spacetime, given its central role in standard cosmology \cite{A&S}. A major breakthrough occurred when the pioneering works in Refs. \cite{APS1,APS2} showed that the initial singularity is replaced, in certain states of LQC, with a phenomenon known as the big bounce, offering a quantum alternative to the classical big bang.

Black holes are one of the most active and prominent topics in GR. The Schwarzschild metric is the simplest classical black hole solution. It represents a nonrotating, electrically neutral, spherical black hole. When external observers cross the event horizon of the Schwarzschild solution, they find that the roles of space and time are reversed. Consequently, the spherically symmetric metric transitions from a static form outside the black hole to a homogeneous but anisotropic one inside. The spacetime in the interior can be interpreted to belong to the Kantowski-Sachs family \cite{K&S,E.Weber}. Building on the success of LQC in addressing cosmological models, researchers within the LQG community have extended the application of loop techniques to the study of black holes (see e.g. Refs. \cite{A&B,LBH1,LBH2,B&V,Chiou,LBH3,LBH4,LBH5,JPS,LBH6,LBH7,LBH8,LBH9,LBH10,LBH11,LBH12,BGMS,LM,GOP,KLEPW,MP,HKSW}). In this context, the Kantowski-Sachs spacetime provides an interesting arena for exploring the quantum nature of black hole interiors. Using this approach, several works have achieved an effective description \cite{AOS,AOS2,AO} and a complete quantization of the Schwarzschild interior employing LQC methods \cite{BH_Con,BH_GAB,BH_Cong,GBA}. In particular, Refs. \cite{BH_GAB,GBA} introduced an extended Hamiltonian formulation as the fundamental distinction necessary for supporting the quantization. Although this extension initially appears to complicate the problem, it actually simplifies the quantum treatment. Just as in FLRW cosmology, applying loop techniques to black hole interiors has led to significant results. One of the most notable is the resolution of the essential singularity, which is replaced by a black hole to white hole transition surface \cite{AOS,AOS2,AO}.

A natural next step in the quantum study of black holes is the inclusion of perturbations. Perturbed black holes play a crucial role in physics. For example, their properties provide insight into the final stages of black hole mergers. Developing a quantum theory capable of describing these perturbations would mark significant progress in the study of quasinormal modes and gravitational wave physics. Despite quantum techniques have been applied to other cosmological models with perturbations (see Refs. \cite{AAN1,AAN2,Ivan,FMO,hyb1,hyb2,hyb-review,hyb-others,AshNe,LMM} for discussions on FLRW models and Refs. \cite{IJV,LAG,KBP,Z&W} for other cosmologies), their application to the Kantowski-Sachs model was an open problem until recently. For spherically symmetric spacetimes like this one, perturbations are typically divided into axial and polar components. At first order in perturbation theory, these two types decouple, allowing for their independent study. While recent efforts have focused on the quantization of axial perturbations \cite{MM}, a complete description including polar perturbations has not been developed yet.

To gain a global understanding of Kantowski-Sachs perturbations, this work extends the analysis initiated in Ref. \cite{MM} by examining polar perturbations. To keep the discussion general, we include a massless scalar field, treating it as a perturbation. However, to maintain our focus on black hole spacetimes, we set its homogeneous contribution to zero. In the absence of this scalar field, our results reduce to those for perturbations of the Schwarzschild interior geometry. Throughout our discussion, we will decompose polar perturbations into spherical harmonics and into Fourier modes in (what in Kantowski-Sachs spacetime is) the radial direction. This decomposition enables spatial integration of the main expressions, allowing us to work directly with the expansion coefficients, which serve as the main variables in the phase space of the perturbative degrees of freedom. Further refinement is achieved via a series of background-dependent and mode-dependent canonical transformations, which, along with an appropriate redefinition of the Lagrange multipliers, lead to the final form of our Hamiltonian formulation. To preserve the canonical structure of the system, a highly convenient property for the quantization of the model, we also redefine the background variables, incorporating quadratic perturbative corrections. Ultimately, the perturbative canonical variables are either gauge invariants or pure gauge. The latter decouple from the physical degrees of freedom and are irrelevant to the main discussion. Beyond Ref. \cite{MM}, key references contributing to our study, which have explored gauge invariant Hamiltonian formulations for different spacetimes, are Refs. \cite{Lan,LMM,DC}. Whilst we follow similar steps, the polar case requires careful treatment, as it is much more complex than its axial counterpart.

Without adopting any fixation of perturbed gauge, the physical dynamics of the perturbations can be generated by a Hamiltonian that only depends on the gauge invariant variables. Under fairly broad conditions, the dynamics of these gauge invariants admit a unique unitary family of Fock quantizations. The demonstration follows from the results presented in Refs. \cite{AT, AT2}. Furthermore, by applying the criteria outlined in Refs. \cite{AT2, BGTT} to this scenario, a well-motivated vacuum definition for the perturbative gauge invariants is obtained. Combined with a quantization of the background, this result paves the way for a hybrid quantization of the model. In particular, the hybrid loop quantum cosmology (hybrid-LQC) formalism provides a robust framework for describing this quantum system with perturbations. This formalism operates under the assumption that, for the quantum geometry, the background plays a dominant role, while perturbations have a secondary effect. The core idea behind this approach is that background and perturbative degrees of freedom can be quantized using different methods, interrelated by the constraints of the system. Our choice of quantum representation is guided by fundamental properties, such as a well-defined representation of background quantities as properly established operators, the resolution of the interior singularity, and a behavior that closely resembles quantum field theory in curved spacetime in effective regimes (or when the background is treated as fixed). These properties hold if we use LQC to incorporate quantum geometric effects into the background, while employing a Fock representation for the perturbative degrees of freedom. For more details, a review of the hybrid-LQC approach can be found in Ref. \cite{hyb-review}, with successful applications in cosmological contexts, e.g., in Refs. \cite{FMO,hyb1,hyb2}.

The article is organized as follows. In Sec. \ref{sec: II}, we introduce the fundamental formulas for the Hamiltonian formulation of the entire system, covering both the perturbed and unperturbed cases. Section \ref{sec: III} is dedicated to deriving the canonical pairs, constraints, and Hamiltonian for polar perturbations. In Sec. \ref{sec: IV}, we refine the perturbative approach while keeping the canonical structure. We also provide a complete set of perturbative gauge invariants and formulate an appropriate Hamiltonian for them. Section \ref{sec: V} outlines the key concepts for quantizing the system using a hybrid approach. In regimes where the background can be described effectively within this approach, a criterion to determine a privileged vacuum state for the perturbations is proposed in Sec. \ref{sec: VI}. Lastly, in Sec. \ref{sec: VII}, we present and discuss the implications of our results. To keep the main text clear and concise, lengthy expressions have been moved to two appendixes. Throughout the paper, we use geometric natural units, where the speed of light, the reduced Planck constant, and the gravitational constant are set to unity.

\section{Hamiltonian framework for Kantowski-Sachs spacetime \label{sec: II}}

In this section, we summarize the key formulas for the Hamiltonian description of a perturbed Kantowski-Sachs spacetime in the presence of a (perturbative) massless scalar field. This overview includes the background framework and the perturbative formulation. We introduce the fundamental concepts, notation, and types of variables necessary for our analysis. A more comprehensive treatment can be found in Refs. \cite{Thiemann,DC,DB3,DB2,DB1}.

\subsection{Background formulation \label{ssec: IIA}}

The Kantowski-Sachs spacetime is a homogeneous but anisotropic solution to Einstein's equations. It can be used to describe  the interior geometry of a Schwarzschild black hole. Its corresponding line element can be written as
\begin{equation}
    \text{d}s^2 = \frac{p_b^2(t)}{L_o^2} \left( -|p_c(t)|\Tilde{N}^2(t) \text{d}t^2 + \frac{1}{|p_c(t)|} \text{d}x^2 \right)+ |p_c(t)|(\text{d}\theta^2 + \sin^2\theta\text{d}\phi^2). 
\end{equation}
To derive this expression, we assume compact spatial sections of the form $\sigma_o = S_o^1 \times S^2$, where $S_o^1$ denotes a 1-sphere with period $L_o$, and $S^2$ is a standard 2-sphere. This compactness assumption simplifies calculations and helps avoid infrared issues (related to spatial sections of infinite volume). In this setting, the coordinate $x$ has a periodicity of $L_o$, while $\theta$ and $\phi$ parametrize the 2-sphere. A noncompact scenario, where $L_o$ grows large, requires a more careful treatment. Interpreting the remaining terms in the expression is less straightforward within the standard Hamiltonian formulation of GR. This complexity stems from two key differences, namely the use of a densitized lapse function, $\Tilde{N}$, instead of the usual lapse function, $N$, and the choice of the metric functions $p_b$ and $p_c$ that are naturally adapted to a triad-connection formulation. 

At the classical level, these differences do not affect physical predictions, allowing for flexibility in their choice. On the one hand, the lapse functions are related by $\Tilde{N} = 4\pi NL_o/\mathcal{V}$, where $\mathcal{V}$ represents the volume of the spatial sections. On the other, the triad-connection formulation for the Kantowski-Sachs spacetime is built upon the densitized triad and its Ashtekar-Barbero $SU(2)$ connection, defined as
\begin{equation}
    E = p_c \text{sin}\theta \tau_3\partial_x + \frac{p_b}{L_o} \text{sin}\theta \tau_2\partial_{\theta} - \frac{p_b}{L_o} \tau_1\partial_{\phi}, \qquad
    A = \frac{c}{L_o}\tau_3\text{d}x + b\tau_2\text{d}\theta - b\text{sin}\theta \tau_1\text{d}\phi + \text{cos}\theta\tau_3\text{d}\phi,
\end{equation}
where $b$ and $c$ are two dynamical variables and $\tau_i$ (with $i=1,2,3$) are the standard basis elements in $\mathfrak{su}(2)$, satisfying the commutation relations $[\tau_i,\tau_j]=\tensor{\epsilon}{_{ij}^k}\tau_k$. We select to work with triad-connection variables beforehand because they will be convenient for our later quantum discussion. However, we emphasize that prequantization results remain independent of adopting this choice and can be equally obtained using other canonical pairs.

We now turn to the Hamiltonian formulation. After a spatial integration, the background action takes the form
\begin{equation}
    \label{eq: IIA-2.3}
    S_0 = \int \bigg( \frac{1}{\gamma}p_b\text{d}b + \frac{1}{2\gamma}p_c\text{d}c - \Tilde{H}_{\text{KS}}[\Tilde{N}]\text{d}t \bigg).
\end{equation}
The connection variables $b$ and $c$ are thus canonically conjugate to the triad variables $p_b$ and $p_c$, with nonvanishing Poisson brackets\footnote{The label $\text{B}$ indicates Poisson brackets of background variables.} defined as $\{b,p_b\}_{\text{B}}=\gamma$ and $\{c,p_c\}_{\text{B}}=2\gamma$, where $\gamma$ is the so-called Immirzi parameter \cite{A&L,Thiemann}. The system evolves both in the $b$-sector and the $c$-sector of the associated phase space, which can be regarded as describing (the conformal factors accompanying) the radial and angular parts of the spatial metric, respectively. Additionally, the background description excludes contributions from the massless scalar field, which is treated as a perturbation with a vanishing homogeneous part. The Hamiltonian constraint, which governs the background dynamics, is
\begin{equation}
    \Tilde{H}_{\text{KS}}[\Tilde{N}] = -\Tilde{N}\frac{L_o}{2}\bigg[ \Omega_b^2 + \frac{p_b^2}{L_o^2} + 2\Omega_b\Omega_c \bigg],
\end{equation}
where $\Omega_b\!=\!bp_b/(\gamma L_o)$ and $\Omega_c\!=\!cp_c/(\gamma L_o)$ are the generators (up to a multiplicative constant) of dilation in the $b$-sector and the $c$-sector. The evolution of any background-dependent quantity can be found computing its Poisson brackets with the Hamiltonian constraint. 

Although our analysis does not require fixing $\Tilde{N}$, setting its inverse equal to $\Omega_b$ allows for a separation of the Hamiltonian into two independent components for the $b$-sector and the $c$-sector. This separation reinforces the interpretation of the selected variables as governing the radial and angular dynamics, providing a physical justification and a natural motivation for their choice. Furthermore, this approach recovers the results of Refs. \cite{A&B, AOS}, confirming its consistency with previous investigations.

\subsection{Perturbative formulation \label{ssec: IIB}}

The perturbed Kantowski-Sachs spacetime is a significantly more complex system to study. That is why first-order perturbations are essential, as they allow us to explore new results in the simplest possible way. To describe these perturbations, we can use their behavior under rotations, given the spherical symmetry of our background. Consider, for instance, a perturbative symmetric spatial tensor, described as a two-form $T$. We can decompose it as
\begin{equation}
    \label{eq: IIB-2.5}
    T = T_{xx}\text{d}x^2 + 2T_{xA}\text{d}x\text{d}x^A + T_{AB}\text{d}x^A\text{d}x^B,
\end{equation}  
where uppercase letters denote indices on the 2-sphere. Under rotations of $S^2$, $T_{xx}$ transforms as a scalar, $T_{xA}$ as a covector, and $T_{AB}$ as a symmetric 2-tensor, with all three components being scalars on $S_o^1$. This type of decomposition applies naturally to other first-order perturbations with different tensorial ranks.

Moreover, using this decomposition, we can fully exploit the symmetries of the Kantowski-Sachs spacetime, including the additional Killing field. More specifically, these symmetries enable us to expand each component in Eq. \eqref{eq: IIB-2.5} into real spherical harmonics and real Fourier modes. Taking $T_{xx}$ as an example, it can be expanded as
\begin{equation}
    T_{xx} = \sum_{\mathfrak{N}_0,\lambda} T^{\mathfrak{n},\lambda}(t) Y_l^m(\theta,\phi)Q_{n,\lambda}(x),
\end{equation}
where the spherical harmonics $Y_l^m$ satisfy $(\square+l(l+1))Y_l^m=0$, and the Fourier modes $Q_{n,\lambda}$ consist of sine $(\lambda\!=\!-)$ and cosine $(\lambda\!=\!+)$ functions with frequency $\omega_n \!=\!2\pi n/L_o$, normalized on the one-sphere.\footnote{In more detail, $Y_l^0$ is the standard spherical harmonic with labels $l$ and $m=0$, while $Y_l^m$ for $m=\pm|m|$ are $\sqrt{2}$ times the real and imaginary parts of the standard harmonic with labels $l$ and $|m|$, multiplied by $(-1)^m$ if $m<0$. In addition, in the case of the Fourier mode with $n=0$, the label $\lambda$ ceases to adopt two values, and $Q_0(x)$ is proportional to the unit function.} To adopt a more compact notation, we have defined the set $\mathfrak{N}_k = \{ (n,l,m) \,|\, n\in\mathbb{N}_0,l\in\{k,k+1,k+2,\cdots\},m\in\{-l,\cdots l\} \}$ for $k\geq 0$, where $\mathbb{N}_0$ denotes the set of nonnegative integers. The mode labels have been written as $\mathfrak{n}=(n,l,m)$. The factors $T^{\mathfrak{n},\lambda}$ are real expansion coefficients that capture the essential dynamical information about $T_{xx}$. This expansion procedure can also be applied to other components of $T$, although it may require spherical harmonics of higher tensorial ranks. In brief, tensor harmonics can be viewed as linear combinations of the metric on $S^2$ and/or its covariant derivative acting on $Y_l^m$. Further details on the construction, normalization, and properties of these harmonics can be found in Chap. 6 of Ref. \cite{DC} and in Ref. \cite{MM}.

Since every perturbation is a scalar on $S^1_o$, the tensorial rank on the 2-sphere is sufficient to classify Kantowski-Sachs perturbations. As the tensorial rank increases, the decomposition becomes more intricate. However, contributions can always be split into two types: axial, $\mathcal{A}$, and polar, $\mathcal{P}$. This distinction is based on how harmonics behave under parity transformations $(\theta,\phi) \mapsto (\pi-\theta,\pi+\phi)$. To be precise, the axial components transform as $\mathcal{A} \rightarrow (-1)^{l+1} \mathcal{A}$, while the polar components transform as $\mathcal{P} \rightarrow (-1)^l \mathcal{P}$. At the perturbative order that we are considering, axial and polar perturbations are decoupled, meaning they can be studied independently. 

The dynamics of Kantowski-Sachs perturbations are treated by following a similar approach to that used for the background. First, starting with the action of the system, we expand its expression in powers of a dimensionless parameter, $\epsilon$, which tracks the perturbation order in a hierarchy. The zeroth-order term corresponds to the background action introduced in Eq. \eqref{eq: IIA-2.3}, while the first-order term vanishes upon integrating over $\sigma_o$. Likewise, contributions from pure second-order terms are discarded. This simplification enables us to focus on the leading-order perturbations using only an effective second-order contribution, without involving higher-order terms. The resulting action of the system, truncated at the quadratic order, is given by
\begin{equation}
    \label{eq: IIB-2.7}
    S = S_0 + \frac{\epsilon^2}{2}\Delta^2_1[S] + O(\epsilon^3).
\end{equation}
The notation $\Delta^2_1$ is used to indicate that a second-order expression is composed of first-order perturbations. In particular, the second-order action can be expressed in the Hamiltonian framework as
\begin{equation}
    \label{eq: IIB-2.8}
    \frac{1}{2}\Delta^2_1[S] = \frac{1}{\kappa} \int_{\mathbb{R}}\text{d}t\int_{\sigma_o}\text{d}^3x \bigg( h_{ab,t}p^{ab} + \varphi_{,t}p - C\Delta[\mathcal{H}] - B^a\Delta[\mathcal{H}_a] - \frac{N}{2}\Delta^2_1[\mathcal{H}] \bigg) .
\end{equation}
Here, $\kappa$ is a constant equal to $16\pi$ in our chosen units, and the comma denotes derivatives. The pair $(h_{ab},p^{ab})$, with lowercase letters indicating spatial indices, corresponds to the first-order perturbations of the spatial metric and their momenta. In turn, the massless scalar field, treated as a perturbation, is described by $(\varphi,p)$ at the considered first-order. Both pairs provide canonically conjugate perturbative variables. The pairs $(C,B^a)$ and $(\Delta[\mathcal{H}],\Delta[\mathcal{H}_a])$, on the other hand, represent the first-order perturbative Lagrange multipliers and gauge constraints. As mentioned in the Introduction, any perturbative description within GR involves a gauge freedom related to the definition of the perturbations. In the Hamiltonian formalism, this freedom translates into the existence of constraints, which are the generators of the perturbative gauge transformations and ensure that only physical perturbations (perturbative gauge invariants) evolve consistently. Finally, $\Delta_1^2[\mathcal{H}]$ constitutes the leading perturbative contribution to the total Hamiltonian. This Hamiltonian description of the perturbative action is obtained by applying the results of Chap. 4 of Ref. \cite{DC} to our Kantowski-Sachs spacetime. A more detailed analysis, including explicit expressions for the constraints and the perturbative Hamiltonian, can be found in that reference.

\section{Polar perturbations \label{sec: III}}

We can further develop the perturbative study of our system, which is governed by action \eqref{eq: IIB-2.8}. Actually, previous works have already analyzed the axial contribution (see Ref. \cite{MM}), which is the simplest to handle. Here, we complete the perturbative description by focusing on the remaining unexplored contribution, namely that of the polar sector. Specifically, in this section we first identify canonical pairs of perturbative variables for this polar sector, then derive the corresponding constraints and Hamiltonian contribution, and finally present a refined version of the action for such polar perturbations.

The symplectic form on the phase space for first-order perturbations arises from the left-hand side of Eq. \eqref{eq: IIB-2.8}, where the polar contribution to the first-order perturbative metric and its conjugate momentum can be expanded using the techniques explained above as
\begin{equation}
    \label{eq: III-3.1}
    \begin{aligned}
        &\begin{aligned}
            \left[h_{ab}\text{d}x^a\text{d}x^b\right]^{\text{po}} &= \sum_{\mathfrak{N}_0,\lambda}\big[h_6^{\mathfrak{n},\lambda}Y_l^mQ_{n,\lambda}\text{d}x^2 + h_3^{\mathfrak{n},\lambda}\tensor{Y}{_l^m_{AB}}Q_{n,\lambda}\text{d}x^A\text{d}x^B\big]\\ 
            &+ \sum_{\mathfrak{N}_1,\lambda} 2h_5^{\mathfrak{n},\lambda}\tensor{Z}{_l^m_A}Q_{n,\lambda}\text{d}x\text{d}x^A + \sum_{\mathfrak{N}_2,\lambda} h_4^{\mathfrak{n},\lambda}\tensor{Z}{_l^m_{AB}}Q_{n,\lambda}\text{d}x^A\text{d}x^B,
        \end{aligned}\\
        &\begin{aligned}
            \bigg[\frac{p_{ab}}{\sqrt{\mathfrak{g}}}\text{d}x^a\text{d}x^b\bigg]^{\text{po}} &= \sum_{\mathfrak{N}_0,\lambda} \bigg[\frac{p_b^4}{L_o^4p_c^2}\Tilde{p}_6^{\mathfrak{n},\lambda}Y_l^mQ_{n,\lambda}\text{d}x^2 + p_c^2\Tilde{p}_3^{\mathfrak{n},\lambda}\tensor{Y}{_l^m_{AB}}Q_{n,\lambda} \text{d}x^A\text{d}x^B\bigg]\\ 
            &+ \sum_{\mathfrak{N}_1,\lambda} 2\Tilde{p}_5^{\mathfrak{n},\lambda} \tensor{Z}{_l^m_A}Q_{n,\lambda}\text{d}x\text{d}x^A + \sum_{\mathfrak{N}_2,\lambda} p_c^2\Tilde{p}_4^{\mathfrak{n},\lambda}\tensor{Z}{_l^m_{AB}}Q_{n,\lambda}\text{d}x^A\text{d}x^B.
        \end{aligned}
    \end{aligned}
\end{equation}
Here, the superscript $\text{po}$ denotes the polar part, while $\mathfrak{g}$ represents the determinant of the spatial metric. The terms $\tensor{Z}{_l^m_A}$, $\tensor{Z}{_l^m_{AB}}$, and $\tensor{Y}{_l^m_{AB}}$ correspond to polar tensorial Regge-Wheeler-Zerilli harmonics of ranks one and two. The expansion coefficients are labeled with subscripts ranging from 3 to 6 (whereas in Ref. \cite{MM} subscripts 1 and 2 were used for the axial degrees of freedom). The first-order contribution to the scalar field and its conjugate momentum,
\begin{equation}
    \label{eq: III-3.2}
    \varphi = \sum_{\mathfrak{N}_0,\lambda} h_7^{\mathfrak{n},\lambda}Y_l^mQ_{n,\lambda}, \qquad \frac{p}{\sqrt{\mathfrak{g}}} = \sum_{\mathfrak{N}_0,\lambda} \Tilde{p}_7^{\mathfrak{n},\lambda}Y_l^mQ_{n,\lambda},
\end{equation}
has only a polar contribution; therefore, the superscript $\text{po}$ is unnecessary. The coefficients in Eqs. \eqref{eq: III-3.1} and \eqref{eq: III-3.2} contain all the relevant dynamical information about the perturbations and serve as phase space variables in the perturbative Hamiltonian framework. Upon redefining the momentum variables in a mode-dependent and background-dependent manner,
\begin{equation}
    p_3^{\mathfrak{n},\lambda} = \frac{\mathcal{V}\Tilde{p}_3^{\mathfrak{n},\lambda}}{2\pi L_o},\qquad p_4^{\mathfrak{n},\lambda} = \frac{(l+2)!}{(l-2)!}\frac{\mathcal{V}\Tilde{p}_4^{\mathfrak{n},\lambda}}{8\pi L_o},\qquad p_5^{\mathfrak{n},\lambda} = l(l+1)\frac{L_o^2}{p_b^2}\frac{\mathcal{V}\Tilde{p}_5^{\mathfrak{n},\lambda}}{2\pi L_o},\qquad p_6^{\mathfrak{n},\lambda} = \frac{\mathcal{V}\Tilde{p}_6^{\mathfrak{n},\lambda}}{4\pi L_o}, \qquad  p_7^{\mathfrak{n},\lambda} = \frac{\mathcal{V}\Tilde{p}_7^{\mathfrak{n},\lambda}}{4\pi L_o},
\end{equation}
we obtain five canonical pairs $(h_i^{\mathfrak{n},\lambda},p_i^{\mathfrak{n},\lambda})$, with Poisson brackets\footnote{The label $\text{P}$ indicates Poisson brackets of perturbative modes.} given by $\{h_i^{\mathfrak{n},\lambda},p_{i'}^{\mathfrak{n}',\lambda'}\}_{\text{P}} = \kappa \delta_{ii'}\delta_{nn'}\delta_{ll'}\delta_{mm'}\delta_{\lambda\lambda'}$. Moreover, with this redefinition we ensure that the perturbative modes and the background variables Poisson commute, resulting in a global canonical set, similar to the axial case in Ref. \cite{MM}.

We continue the discussion by addressing the polar contribution to the first-order perturbative gauge constraints. In Eq. \eqref{eq: IIB-2.8}, these terms appear in the central part of the equation. To proceed, we first expand the perturbative Lagrange multipliers as follows
\begin{equation}
    C = - \sum_{\mathfrak{N}_0,\lambda} \kappa\frac{\mathcal{V}}{8\pi L_o} \Tilde{N} f_0^{\mathfrak{n},\lambda}Y_l^mQ_{n,\lambda}, \qquad B_a^{\text{po}}\text{d}x^a = \sum_{\mathfrak{N}_0,\lambda} \kappa\frac{p_b^2}{L_o^2} k_0^{\mathfrak{n},\lambda}Y_l^mQ_{n,\lambda}\text{d}x + \sum_{\mathfrak{N}_1,\lambda} \kappa|p_c|q_0^{\mathfrak{n},\lambda}\tensor{Z}{_l^m_A}Q_{n,\lambda}\text{d}x^A.
\end{equation}
The subscript $0$ labels the expansion coefficients, indicating that they are not true degrees of freedom. Following Ref. \cite{DC}, we define the perturbative constraints for the polar sector as
\begin{equation}
    \label{eq: III-3.5}
    \kappa\mathbf{C}_1[q_0^{\mathfrak{n},\lambda}] = \int_{\sigma_o}\text{d}^3x B_A^{\text{po}}\Delta[\mathcal{H}^A]^{\text{po}}, \qquad \kappa\mathbf{C}_2[k_0^{\mathfrak{n},\lambda}] = \int_{\sigma_o}\text{d}^3x \frac{L_o^2}{p_b^2}|p_c|B_x^{\text{po}}\Delta[\mathcal{H}_x]^{\text{po}}, \qquad \kappa C_3[f_0^{\mathfrak{n},\lambda}] = \int_{\sigma_o}\text{d}^3x C\Delta[\mathcal{H}].
\end{equation}
In this way, we conclude that $C_3$ represents the integrated form of the perturbative Hamiltonian constraint, while $\mathbf{C}_1$ and $\mathbf{C}_2$ correspond to the scalar and vector contributions to the integrated version of the perturbative momentum constraint. To keep the discussion simple, we have included their explicit expressions in Appendix \ref{sec: app-1}. Further calculations reveal that the above constraints do not form an Abelian set. This means that, under Poisson brackets, they do not all commute exactly with each other at the perturbative order that we are considering, but instead generate linear combinations of the constraints. In particular, we find
\begin{equation}
    \begin{aligned}
        &\left\{\mathbf{C}_1[q_0^{\mathfrak{n},\lambda}],\mathbf{C}_2[k_0^{\mathfrak{n},\lambda}]\right\}_{\text{P}} = 0,\\
        &\left\{\mathbf{C}_1[q_0^{\mathfrak{n},\lambda}],C_3[f_0^{\mathfrak{n},\lambda}]\right\}_{\text{P}} =-\sum_{\mathfrak{N}_1,\lambda} 2\kappa\Tilde{N}l(l+1) q_0^{\mathfrak{n},\lambda}f_0^{\mathfrak{n},\lambda}\frac{\Tilde{H}_{\text{KS}}}{L_o},\\
        &\left\{\mathbf{C}_2[k_0^{\mathfrak{n},\lambda}],C_3[f_0^{\mathfrak{n},\lambda}]\right\}_{\text{P}} = -\sum_{\mathfrak{N}_0,\lambda}2\kappa\Tilde{N}\lambda\omega_n|p_c|k_0^{\mathfrak{n},\lambda}f_0^{\mathfrak{n},-\lambda}\frac{\Tilde{H}_{\text{KS}}}{L_o}.
    \end{aligned}
\end{equation}
An Abelianized set of constraints simplifies the Hamiltonian analysis, particularly in its quantum counterpart. We can reach a set of this kind by including in $C_3$ a convenient term proportional to the background Hamiltonian. Although this redefinition does not modify any key expression, it requires an adjustment of the densitized lapse function at second order in the perturbations to ensure consistency in the Hamiltonian expression of the truncated action, given by Eqs. \eqref{eq: IIB-2.7} and \eqref{eq: IIB-2.8}. Specifically, we redefine
\begin{equation}
    \mathbf{C}_3[f_0^{\mathfrak{n},\lambda}] = C_3[f_0^{\mathfrak{n},\lambda}] + \sum_{\mathfrak{N}_0,\lambda}\Tilde{N}f_0^{\mathfrak{n},\lambda}\frac{\Tilde{H}_{\text{KS}}}{L_o|p_c|}\bigg[\frac{L_o^2}{p_b^2}p_c^2h_6^{\mathfrak{n},\lambda}+ 2h_3^{\mathfrak{n},\lambda}\bigg], \quad \hat{N} = \Tilde{N} - \epsilon^2\sum_{\mathfrak{N}_0,\lambda}\Tilde{N}\frac{f_0^{\mathfrak{n},\lambda}}{L_o|p_c|}\bigg[\frac{L_o^2}{p_b^2}p_c^2h_6^{\mathfrak{n},\lambda}+ 2h_3^{\mathfrak{n},\lambda}\bigg].
\end{equation}
With these redefinitions, the truncated action remains unchanged, as the additional terms are of higher order than the truncation \cite{LMM,MM}. Exploiting the freedom in the choice of Lagrange multipliers, we introduce perturbative corrections in their redefinitions, a strategy that will be repeatedly employed in the next sections to rewrite the action. Since this is the first example, we have chosen to include it explicitly in the main text, while any subsequent redefinitions will be referenced in the corresponding appendix or briefly commented.

The dynamics of the polar modes is governed by the perturbative Hamiltonian, which is derived from Eq. \eqref{eq: IIB-2.8} by spatially integrating $\Delta_1^2[\mathcal{H}]^{\text{po}}$. The integrated expression, denoted $\Tilde{H}^{\text{po}}$, is presented in Appendix \ref{sec: app-1}. During its derivation, we choose to adjust the density weight of the Hamiltonian to account for the densitized lapse function, simplifying the connection between the perturbative results and the background expressions \cite{BH_Cong,GBA}. 

By incorporating all these calculations in Eq. \eqref{eq: IIB-2.8}, we obtain a much simpler form for the action of the polar perturbations. The second-order action becomes
\begin{equation}
    \label{eq: III-3.8}
    \begin{aligned}
        \frac{1}{2}\Delta^2_1[S]^{\text{po}} &= \frac{1}{\kappa}\int\Bigg(\sum_{\mathfrak{N}_0,\lambda} \left[p_3^{\mathfrak{n},\lambda}\text{d}h_3^{\mathfrak{n},\lambda} + p_6^{\mathfrak{n},\lambda}\text{d}h_6^{\mathfrak{n},\lambda} + p_7^{\mathfrak{n},\lambda}\text{d}h_7^{\mathfrak{n},\lambda}\right] + \sum_{\mathfrak{N}_1,\lambda} p_5^{\mathfrak{n},\lambda}\text{d}h_5^{\mathfrak{n},\lambda}\\
        &+ \sum_{\mathfrak{N}_2,\lambda} p_4^{\mathfrak{n},\lambda}\text{d}h_4^{\mathfrak{n},\lambda} - \kappa\left[\textbf{C}_1[q_0^{\mathfrak{n},\lambda}] + \textbf{C}_2[k_0^{\mathfrak{n},\lambda}] + \textbf{C}_3[f_0^{\mathfrak{n},\lambda}] + \Tilde{H}^{\text{po}}[\hat{N}]\right]\text{d}t\Bigg).
    \end{aligned}
\end{equation}

\section{Perturbative gauge invariants \label{sec: IV}}

The results of the previous section provide all the necessary ingredients for a Hamiltonian formulation of the first-order polar perturbations. At this stage, we can compute dynamical equations and study the evolution of the modes. However, their physical interpretation remains unclear since our current formulation does not yet incorporate first-order perturbative gauge invariants. In the Hamiltonian framework, these invariants are quantities that Poisson commute with all the perturbative constraints (treating the background as fixed). Once identified, their dynamics can yield meaningful physical results.

This section introduces a systematic method within the Hamiltonian formalism to identify first-order perturbative gauge invariants. Its validity is supported by successful applications to other cosmological scenarios \cite{Lan,LMM,DC} and to Kantowski-Sachs axial perturbations \cite{MM}. The main idea behind this approach is that, under fairly general conditions, any set of canonical pairs provides a valid description of the system. By applying a sequence of appropriate canonical transformations, we incorporate the perturbative constraints into the set of canonical variables while redefining the remaining terms to preserve the canonical Poisson bracket structure. As a result, the final set consists of perturbative gauge invariants and perturbative constraints (which are also gauge invariant, as they vanish in any gauge), along with their respective canonical momenta. Notably, this proposal does not require fixing the perturbative gauge at any stage to obtain the final results. Although extending this method from the axial to the polar case may seem natural, its significance lies in the nontrivial verification that such canonical transformations exist and enable us to cast the Hamiltonian description into this particular form. Since the perturbed scalar field has no homogeneous contribution, its perturbations decouple from the geometric perturbations, forming our first gauge invariant pair. Identifying the remaining invariant pair, however, is less straightforward and requires the use of the aforementioned systematic method. To streamline the analysis, we focus exclusively on modes with $l\geq2$, allowing us to address all perturbative modes at once and simplifying many of the forthcoming equations. Later, we will briefly discuss the excluded modes with $l=0$ or $l=1$. Unless otherwise noted, we follow the same notation as in the previous section for the perturbative constraints and the Hamiltonian.

We will divide the analysis that allows us to identify the second gauge invariant pair into intermediate steps. For now, we focus on the first two perturbative constraints in Eq. \eqref{eq: III-3.5}, which can be addressed by introducing a background-dependent and mode-dependent canonical transformation. The generator of this transformation is a type-$3$ function, defined as
\begin{equation}
    \label{eq: IV-4.1}
    \begin{aligned}
        \mathscr{F}_{(3)}[\Bar{h}_i^{\mathfrak{n},\lambda},p_i^{\mathfrak{n},\lambda}] &= -\sum_{\mathfrak{N}_2,\lambda} \bigg[ |p_c|\Bar{h}_3^{\mathfrak{n},\lambda}p_3^{\mathfrak{n},\lambda} + |p_c|\Bar{h}_5^{\mathfrak{n},\lambda}p_4^{\mathfrak{n},\lambda} + \frac{p_b^2}{L_o^2}\Bar{h}_6^{\mathfrak{n},\lambda}p_5^{\mathfrak{n},\lambda} + \frac{p_b^2}{L_o^2|p_c|}\Bar{h}_4^{\mathfrak{n},\lambda}p_6^{\mathfrak{n},\lambda} + \Bar{h}_7^{\mathfrak{n},\lambda}p_7^{\mathfrak{n},\lambda}\\
        &+ \lambda\omega_n\frac{|p_c|}{2}p_5^{\mathfrak{n},\lambda}\Bar{h}_5^{\mathfrak{n},-\lambda} - 2\lambda\omega_n|p_c|\bigg(\Omega_b\Bar{h}_4^{\mathfrak{n},\lambda} - (\Omega_b+\Omega_c)\Bar{h}_3^{\mathfrak{n},\lambda} -  \frac{p_b^2}{L_o^2|p_c|}p_6^{\mathfrak{n},\lambda}\bigg)\Bar{h}_6^{\mathfrak{n},-\lambda}\\
        &- l(l+1)\Omega_b\Bar{h}_5^{\mathfrak{n},\lambda}\Bar{h}_4^{\mathfrak{n},\lambda} - \frac{l(l+1)}{2}\frac{L_o^2}{p_b^2}\bigg(\omega_n^2p_c^2\Omega_b+\frac{1}{2}(l+2)(l-1)\frac{p_b^2}{L_o^2}(\Omega_b+\Omega_c)\bigg)[\Bar{h}_5^{\mathfrak{n},\lambda}]^2\\
        &- l(l+1)\frac{|p_c|}{2}\Bar{h}_5^{\mathfrak{n},\lambda}p_3^{\mathfrak{n},\lambda} - 2\left(\omega_n^2p_c^2\Omega_b+\frac{1}{2}l(l+1)\frac{p_b^2}{L_o^2}(\Omega_b+\Omega_c)\right)[\Bar{h}_6^{\mathfrak{n},\lambda}]^2\bigg].
    \end{aligned}
\end{equation}
As usual, the relationship between the old and new perturbative variables is given by
\begin{equation}
    \bar{p}_i^{\mathfrak{n},\lambda}=-\frac{\partial\mathscr{F}_{(3)}}{\partial\ \bar{h}_i^{\mathfrak{n},\lambda}}, \qquad h_i^{\mathfrak{n},\lambda}=-\frac{\partial\mathscr{F}_{(3)}}{\partial p_i^{\mathfrak{n},\lambda}}.  
\end{equation}

In this canonical transformation, as well as in future ones, the background terms are treated in practice as fixed contributions. At first glance, this might seem contradictory, given that one of the main goals of this work is to develop a Hamiltonian formulation that treats dynamically background and perturbations altogether. Moreover, by adopting this procedure, not only do we treat the background as nondynamical, but we also risk compromising the global canonical structure of the system, which we aim to preserve as it is essential for implementing a hybrid quantization. The consistency of this approach lies in the fact that, with each canonical transformation for the perturbations, the background variables can be adjusted accordingly. More precisely, they can be corrected with quadratic perturbative terms. Note that this is possible precisely because the zero modes of the model are treated exactly up to the order of our perturbative truncation in the action. For instance, in the previous transformation, the corrected background variables, denoted with overbar notation, satisfy
\begin{equation}
    \frac{1}{\gamma}\Bar{p}_b\text{d}\Bar{b} + \frac{1}{2\gamma}\Bar{p}_c\text{d}\Bar{c} = \frac{1}{\gamma}p_b\text{d}b + \frac{1}{2\gamma}p_c\text{d}c - \frac{\epsilon^2}{\kappa}\text{d}\mathscr{F}\big|_{\text{B}}.
\end{equation}
In the last term on the right-hand side, a special notation has been used to emphasize that the exterior derivative acts only on the background variables\footnote{\label{note2} Additionally, we have omitted the number indicating the type of generating function. For this discussion, the type is not critical, as the differences between generating functions involve only purely perturbative terms. This criterion will be applied to other formulas as well.}. The explicit expressions for the corrected background terms are not necessary for the current discussion, as they do not affect any significant results at the perturbative order we are working with. In other words, the perturbative gauge constraints and Hamiltonian remain the same (at our truncation order in the action) whether we use the corrected or noncorrected background variables. Therefore, although the background undergoes corrections with each canonical transformation of the perturbations, we will retain its original notation. More details regarding the quadratic corrections can be found in the appendix of Ref. \cite{MM} or in Ref. \cite{LMM}.

In terms of the new set of canonical variables the first two perturbative constraints are greatly simplified to 
\begin{equation}
    \mathbf{C}_1[q_0^{\mathfrak{n},\lambda}] = \sum_{\mathfrak{N}_2,\lambda} 2q_0^{\mathfrak{n},\lambda}\Bar{p}_5^{\mathfrak{n},\lambda}, \qquad \mathbf{C}_2[k_0^{\mathfrak{n},\lambda}] = \sum_{\mathfrak{N}_2,\lambda}k_0^{\mathfrak{n},\lambda}\Bar{p}_6^{\mathfrak{n},\lambda},
\end{equation}
while the third constraint, after an appropriate redefinition listed in Appendix \ref{sec: app-2}, becomes
\begin{equation}
    \begin{aligned}
        \Tilde{\mathbf{C}}_3[f_0^{\mathfrak{n},\lambda}] &= -\sum_{\mathfrak{N}_2,\lambda} \Tilde{N}f_0^{\mathfrak{n},\lambda}\bigg[\Omega_c\Bar{p}_4^{\mathfrak{n},\lambda} - \bigg(2\Omega_b\Omega_c + \frac{l^2+l+2}{2}\frac{p_b^2}{L_o^2}\bigg)\Bar{h}_4^{\mathfrak{n},\lambda}\\
        &+ \Omega_b\Bar{p}_3^{\mathfrak{n},\lambda} - \bigg(2\Omega_b(\Omega_b +\Omega_c) + \frac{l^2+l+2}{2}\frac{p_b^2}{L_o^2} + \omega_n^2p_c^2\bigg)\Bar{h}_3^{\mathfrak{n},\lambda}\bigg].
    \end{aligned}
\end{equation}
Owing to the background dependence of the canonical transformation, the perturbative Hamiltonian acquires a correction, analogous to the one that arises in any Hamiltonian when a time-dependent canonical transformation is considered. This correction can be computed from the generating function. Specifically, the new perturbative Hamiltonian can be expressed as $\kappa K^{\text{po}}=\kappa\Tilde{H}^{\text{po}}+\{\mathscr{F},\Tilde{H}_{\text{KS}}\}_{\text{B}}$, where, as mentioned before, we use a special notation to indicate that the Poisson brackets act only on the background variables and we omit the subscript from the generating function (see footnote \ref{note2}). Moreover, to simplify our expressions, from now on we will use a prime after a phase space function to denote its Poisson bracket with $\Tilde{H}_{\text{KS}}$ with respect to the background variables, e.g. $\mathscr{F}' = \{\mathscr{F},\Tilde{H}_{\text{KS}}\}_{\text{B}}$.

Once the corrected Hamiltonian is computed, it can be divided into three components, which we now describe. The first block contains terms proportional to the background Hamiltonian, $\Tilde{H}_{\text{KS}}$. The second block consists of contributions proportional to the first two perturbative constraints, namely $\Bar{p}_5^{\mathfrak{n},\lambda}$ and $\Bar{p}_6^{\mathfrak{n},\lambda}$. The third block includes the remaining terms. The first two components do not affect the dynamics and can be removed (at our perturbative truncation order in the action) by redefining the lapse function, $\bar{N}$, and the Lagrange multipliers associated with the first two constraints, $\mathbf{q}_0^{\mathfrak{n},\lambda}$ and $\mathbf{k}_0^{\mathfrak{n},\lambda}$. In this setup, the third block of $K^{\text{po}}$ is the only part that significantly influences the dynamics and, as a result, is referred to as the true polar perturbative Hamiltonian, denoted as $\bar{H}^{\text{po}}$. Explicit formulas for the above three redefinitions and for the true Hamiltonian expression are provided in Appendix \ref{sec: app-2}.

In the same spirit as before, we introduce another canonical transformation. This time, the goal is to address the third constraint and complete our systematic identification of the perturbative gauge invariants, along the lines explained at the beginning of the section. This second transformation is generated by a type-$3$ function, defined as\footnote{The appearance of $\Omega_b$ in the denominator might pose a challenge at the quantum level, as representing the inverse of $\Omega_b$ is not straightforward. However, this issue arises only in the generating function, and does not affect the relationship between the new and old canonical variables, nor the term correcting the new perturbative Hamiltonian, which are the relevant physical quantities.}
\begin{equation}
    \label{eq: IV-4.6}
    \begin{aligned}
        \mathscr{G}_{(3)}[Q_i^{\mathfrak{n},\lambda},\Bar{p}_i^{\mathfrak{n},\lambda}] &= -\sum_{\mathfrak{N}_2,\lambda}\bigg[ Q_3^{\mathfrak{n},\lambda}\Bar{p}_3^{\mathfrak{n},\lambda} - \frac{l^2+l+2}{4\Omega_b}\frac{p_b^2}{L_o^2}Q_4^{\mathfrak{n},\lambda}\Bar{p}_4^{\mathfrak{n},\lambda} + Q_5^{\mathfrak{n},\lambda}\Bar{p}_5^{\mathfrak{n},\lambda} +
        Q_6^{\mathfrak{n},\lambda}\Bar{p}_6^{\mathfrak{n},\lambda} + 
        Q_7^{\mathfrak{n},\lambda}\Bar{p}_7^{\mathfrak{n},\lambda} + \Omega_bQ_4^{\mathfrak{n},\lambda}\Bar{p}_3^{\mathfrak{n},\lambda}\\
        &+ \bigg(\frac{l^2+l+2}{4\Omega_b}\frac{p_b^2}{L_o^2}\bigg[2\Omega_b\Omega_c+\frac{l^2+l+2}{2}\frac{p_b^2}{L_o^2}\bigg] - \Omega_b\bigg[2\Omega_b(\Omega_b+\Omega_c)+\frac{l^2+l+2}{2}\frac{p_b^2}{L_o^2}+\omega_n^2p_c^2\bigg]\bigg)\\
        &\times\frac{[Q_4^{\mathfrak{n},\lambda}]^2}{2} + \frac{[\bar{p}_4^{\mathfrak{n},\lambda}]^2}{4\Omega_b} - (\Omega_b+\Omega_c)[Q_3^{\mathfrak{n},\lambda}]^2 - \bigg(2\Omega_b(\Omega_b+\Omega_c)+\frac{l^2+l+2}{2}\frac{p_b^2}{L_o^2}+\omega_n^2p_c^2\bigg)Q_3^{\mathfrak{n},\lambda}Q_4^{\mathfrak{n},\lambda} \bigg].
    \end{aligned}
\end{equation}
Once again, the relationships between the perturbative variables can be computed using
\begin{equation}
    P_i^{\mathfrak{n},\lambda} = -\frac{\partial \mathscr{G}_{(3)}}{\partial Q_i^{\mathfrak{n},\lambda}}, \qquad \bar{h}_i^{\mathfrak{n},\lambda} = -\frac{\partial \mathscr{G}_{(3)}}{\partial \bar{p}_i^{\mathfrak{n},\lambda}}.
\end{equation}

Under this transformation, the first two perturbative constraints remain unchanged, but the third one is significantly simplified. Their new expressions are
\begin{equation}
     \mathbf{C}_1[q_0^{\mathfrak{n},\lambda}] = \sum_{\mathfrak{N}_2,\lambda} 2q_0^{\mathfrak{n},\lambda}P_5^{\mathfrak{n},\lambda}, \qquad \mathbf{C}_2[k_0^{\mathfrak{n},\lambda}] = \sum_{\mathfrak{N}_2,\lambda}k_0^{\mathfrak{n},\lambda}P_6^{\mathfrak{n},\lambda}, \qquad \Tilde{\mathbf{C}}_3[f_0^{\mathfrak{n},\lambda}] = -\sum_{\mathfrak{N}_2,\lambda} \Tilde{N}f_0^{\mathfrak{n},\lambda}P_4^{\mathfrak{n},\lambda}.
\end{equation}
Since we are dealing with another background-dependent canonical transformation, the Hamiltonian of the perturbations changes to $\kappa \mathbf{H}^{\text{po}}=\kappa\bar{H}^{\text{po}}+\mathscr{G}'$ and, as previously noted, it can be divided into three blocks. By redefining the lapse function and the Lagrange multiplier associated with the third perturbative constraint, we eliminate the nondynamical content from $\mathbf{H}^{\text{po}}$. Consequently, the resulting true polar perturbative Hamiltonian, $\Tilde{\textbf{H}}^{\text{po}}$, takes a much more manageable form, given by
\begin{equation}
    \label{eq: IV-4.9}
    \kappa\Tilde{\textbf{H}}^{\text{po}}[\Tilde{N}] = \sum_{\mathfrak{N}_2,\lambda}\frac{\Tilde{N}}{2}\bigg[ [P_7^{\mathfrak{n},\lambda}]^2 + \bigg(\omega_n^2p_c^2 + l(l+1)\frac{p_b^2}{L_o^2}\bigg)[Q_7^{\mathfrak{n},\lambda}]^2 + \mathfrak{A}[Q_3^{\mathfrak{n},\lambda}]^2 +  \mathfrak{B}[P_3^{\mathfrak{n},\lambda}]^2 + \mathfrak{C}Q_3^{\mathfrak{n},\lambda}P_3^{\mathfrak{n},\lambda}\bigg].
\end{equation}
In Appendix \ref{sec: app-2}, we give the expressions of the redefined lapse and Lagrange multiplier, now denoted as $\Tilde{\mathbf{N}}$ and $\mathbf{f}_0^{\mathfrak{n},\lambda}$, along with the three background-dependent and mode-dependent coefficients ($\mathfrak{A}$, $\mathfrak{B}$, and $\mathfrak{C}$) that appear in Eq. \eqref{eq: IV-4.9}. 

At this stage, we have identified the polar perturbative gauge invariants for our Kantowski-Sachs model and determined two canonical pairs to describe them \footnote{Verifying that these pairs are perturbative gauge invariants is straightforward, as they commute with the generators of the perturbative gauge transformations by the very construction of our canonical perturbative variables.}, namely $(Q_3^{\mathfrak{n},\lambda},P_3^{\mathfrak{n},\lambda})$ and $(Q_7^{\mathfrak{n},\lambda},P_7^{\mathfrak{n},\lambda})$. The only nontrivial dynamics for the perturbative modes correspond to these invariant pairs. The remaining pairs consist of perturbative constraints and their conjugate variables, which are purely gauge and thus trivial to describe. The Hamiltonian in Eq. \eqref{eq: IV-4.9}, which depends exclusively on the invariant pairs thanks to our approach, allows us to derive the (gauge invariant) equations of motion. However, before proceeding, it is useful to refine the Hamiltonian formulation, especially if we intend to extend our analysis to a quantum treatment. Since any linear (background-dependent) combination of the above invariants is a gauge invariant, we can apply a final canonical transformation that affects only the pair $(Q_3^{\mathfrak{n},\lambda},P_3^{\mathfrak{n},\lambda})$, eliminating the mixed configuration-momentum term in the Hamiltonian, which would otherwise complicate the quantization process. This transformation is generated by a type-$3$ generating function defined as
\begin{equation}
    \mathscr{K}_{(3)}[\Tilde{Q}_i^{\mathfrak{n},\lambda},P_i^{\mathfrak{n},\lambda}] = - \sum_{\mathfrak{N}_2,\lambda}\bigg[ \sum_{j=4}^7 \Tilde{Q}_j^{\mathfrak{n},\lambda}P_j^{\mathfrak{n},\lambda} + \sqrt{\mathfrak{B}}\Tilde{Q}_3^{\mathfrak{n},\lambda}P_3^{\mathfrak{n},\lambda} + \frac{1}{4}\bigg(\mathfrak{C}-\frac{1}{2}\frac{\mathfrak{B}'}{\mathfrak{B}}\bigg)[\Tilde{Q}_3^{\mathfrak{n},\lambda}]^2\bigg],
\end{equation}
where we have used again our compact notation for the Poisson bracket of $\Tilde{H}_{\text{KS}}$ with respect to the background variables. In Appendix \ref{sec: app-2} we show that $\mathfrak{B}$ is positive definite, so that we can ensure that the transformation is well defined\footnote{The fractional term in this expression includes the function $\Omega_b$ and will also appear in the Hamiltonian, requiring careful handling in the next section when addressing quantization.}. The new and old perturbative variables are related via
\begin{equation}
    Q_i^{\mathfrak{n},\lambda} = -\frac{\partial \mathscr{K}_{(3)}}{\partial P_i^{\mathfrak{n},\lambda}}, \qquad \Tilde{P}_i^{\mathfrak{n},\lambda} = -\frac{\partial \mathscr{K}_{(3)}}{\partial \Tilde{Q}_i^{\mathfrak{n},\lambda}},
\end{equation}
and the Hamiltonian, defined as $\kappa\mathscr{H}^{\text{po}} = \kappa\Tilde{\textbf{H}}^{\text{po}} +\mathscr{K}'$, takes on a much more manageable form, given by
\begin{equation}
    \begin{aligned}
        \kappa\mathscr{H}^{\text{po}}[\Tilde{N}] &= \sum_{\mathfrak{N}_2,\lambda}\frac{\Tilde{N}}{2}\bigg[ [\Tilde{P}_7^{\mathfrak{n},\lambda}]^2 + \bigg(\omega_n^2p_c^2 + l(l+1)\frac{p_b^2}{L_o^2}\bigg)[\Tilde{Q}_7^{\mathfrak{n},\lambda}]^2 + [\Tilde{P}_3^{\mathfrak{n},\lambda}]^2\\
        &+ \bigg(\mathfrak{A}\mathfrak{B}-\frac{1}{4}\bigg[\mathfrak{C}-\frac{1}{2}\frac{\mathfrak{B}'}{\mathfrak{B}}\bigg]^2-\frac{1}{4}\bigg[\mathfrak{C}-\frac{1}{2}\frac{\mathfrak{B}'}{\mathfrak{B}}\bigg]'\bigg)[\Tilde{Q}_3^{\mathfrak{n},\lambda}]^2 \bigg].
    \end{aligned}
\end{equation}

To summarize, we incorporate into Eq. \eqref{eq: III-3.8} these three canonical transformations along with all the necessary redefinitions, including those of the lapse function, perturbative constraints, perturbative Hamiltonian, and Lagrange multipliers. The resulting expression for the action of the polar perturbations (with $l\geq 2$) is
\begin{equation}
    \frac{1}{2}\Delta^2_1[S]^{\text{po}}\bigg|_{l\geq 2} = \frac{1}{\kappa}\int\Bigg(\sum_{\mathfrak{N}_2,\lambda}\sum_{j=3}^{7} \Tilde{P}_j^{\mathfrak{n},\lambda}\text{d}\Tilde{Q}_j^{\mathfrak{n},\lambda} - \kappa\left[\textbf{C}_1[\textbf{q}_0^{\mathfrak{n},\lambda}] + \textbf{C}_2[\textbf{k}_0^{\mathfrak{n},\lambda}] + \Tilde{\textbf{C}}_3[\textbf{f}_0^{\mathfrak{n},\lambda}] + \mathscr{H}^{\text{po}}[\Tilde{\textbf{N}}]\right]\text{d}t\Bigg).
\end{equation}
These results, particularly the expressions for the perturbative constraints and Hamiltonian, are fundamental for a consistent quantum treatment of the relevant (gauge invariant) perturbative degrees of freedom. 

The discussion of the perturbative modes with $l<2$ completes the study of the polar perturbations. In this discussion, the Hamiltonian formalism proves to be a highly useful approach, as it allows us to assess the physical relevance of those modes almost straightforwardly. First, just as in the general case, the modes of the perturbative scalar field remain gauge invariant and can be treated in the same way as in the case with $l\geq 2$. Regarding the remaining perturbative modes, namely those arising from the geometry, the analysis differs when $l\leq 1$. Proceeding in descending order, we first examine the case $l=1$. In this situation, not all canonical pairs are present. Specifically, it is easy to see that the pair $(h_4^{\mathfrak{n},\lambda},p_4^{\mathfrak{n},\lambda})$ does not exist when $l=1$. Then, excluding the scalar field contribution, three perturbative pairs remain per mode. On the other hand, a direct evaluation of the expressions for the perturbative constraints in Eq. \eqref{eq: app-A1} shows that these three constraints remain independent and none vanish for $l=1$. Thus, without additional calculations, we can already conclude that no extra physical degrees of freedom exist, because the number of perturbative pairs equals the number of perturbative constraints. Applying a similar analysis to the case with $l=0$, it is straightforward to check that, in addition to $(h_4^{\mathfrak{n},\lambda},p_4^{\mathfrak{n},\lambda})$, the pair $(h_5^{\mathfrak{n},\lambda},p_5^{\mathfrak{n},\lambda})$ also vanishes now. However, there are only two nonzero and independent perturbative gauge constraints in this case. Hence, the number of perturbative pairs again matches the number of perturbative constraints, leading to the same conclusion as before. All in all, the $l<2$ modes (with $\mathfrak{n}\neq 0$) do not correspond to relevant physical degrees of freedom, a fact which justifies their separation from the general discussion. Notably, the zero modes (i.e., $n=l=m=0$) have been discarded in the above analysis. Including them would be redundant, as our formulation already incorporates homogeneous degrees of freedom through the background variables and in principle accounts for the backreaction of the perturbative modes. Moreover, since these modes do not introduce new dynamical degrees of freedom beyond those already present in the background, given that they can be absorbed into it through a proper redefinition, their explicit treatment would not provide additional physical insights.

\section{Hybrid quantization \label{sec: V}}

The results that we have presented so far establish a suitable Hamiltonian framework for the quantum analysis of the background and the polar perturbations in our Kantowski-Sachs model. Given its successful application to other relevant cosmologies, we propose adopting the hybrid-LQC formalism for this analysis \cite{hyb1,hyb-review}. One of its main advantages is that it provides a joint quantum treatment of the entire perturbed system. This formalism relies on the assumption that there exists a regime in which the most significant quantum gravitational effects are primarily concentrated on the background sector (an assumption that appears reasonable in our perturbed model). Another key strength of this formalism is its ability to provide a unitary evolution of the perturbations within the regime of (linear) quantum field theory on a curved spacetime, which is valid when the background behaves effectively (i.e., following the trajectories of an effective Hamiltonian, which in particular may be the classical one of GR) or simply when it is treated as fixed. While other various quantization schemes can indeed be explored starting from our general Hamiltonian description of the background and its perturbations, in the following we base our discussion on reasonable assumptions, such as the quantum resolution of the essential singularity in Kantowski-Sachs and a proper regularization of the background-geometry operators. This quantum resolution of singularities, together with the related regularization of the inverse triad operators, in particular, is necessary to handle the perturbative Hamiltonian, since the background-dependent coefficients that appear in it include inverse powers of the densitized triad variables of the model. The commented assumptions guarantee a consistent and well-motivated quantum treatment and, remarkably, are satisfied in the hybrid-LQC formalism, implemented along the lines discussed in Ref. \cite{MM}.

Therefore, for the background we adopt a discrete triad representation, as it is usually done in LQC. While a more detailed explanation can be found in Refs. \cite{BH_GAB, GBA}, here we summarize the main ideas. Specifically, the LQC quantization of the background that we propose was already employed in Ref. \cite{MM} in the case of axial perturbations (note that this background coincides for the axial and polar sectors). Our choice of LQC representation is primarily motivated by its good properties, as noted above and further explained later in our discussion. However, if an alternative quantum representation were found to yield satisfactory results for the background, or another quantization scheme within LQC proved more successful, one might choose it to replace the specific LQC representation that we are considering within the hybrid formalism. 

With these premises, the quantum dynamics for the background emerge from an effective formulation that incorporates two regularization parameters (one radial and one angular), each associated with a different sector. By extending the phase space of the system to include these parameters (along with suitable conjugate momenta), this effective formulation admits a Hamiltonian description which reproduces the original classical one in the limit where the parameters become negligible. Importantly, within effective trajectories, curvature invariants remain finite, and the essential singularity is resolved through a transition surface. Furthermore, when applying these results to the black hole interior, modifications with respect to GR remain small near the horizon (for small regularization parameters) and allow for a smooth extension of the geometry to the exterior.

Any function of the phase space of the background geometry can be expressed in terms of holonomies and fluxes, which are the fundamental variables in LQG, constructed from connections and triads of the background. To carry out the quantization, we employ the extended phase space formalism mentioned above \cite{AOS2,MM}. An LQC representation is used for the geometric degrees of freedom of the background, while the regularization parameters adopt a Schrödinger representation. A crucial aspect is the use of the MMO prescription (named after its proposers, Mart\'{\i}n-Benito, Mena Marug\'an and Olmedo \cite{MMO}) to handle noncommuting terms. This prescription allows us to define well-behaved quantum operators for any background-dependent function in terms of holonomies and triad operators. This quantization leads to dynamics with superselected subspaces. The ability to define any background function appearing in the Hamiltonian description as a well-defined quantum operator within the superselected Hilbert (sub)space of the model further reinforces the feasibility of the proposed quantization scheme. As a particular result, we can construct a quantum Hamiltonian constraint in the absence of perturbations, for which physical solutions (those annihilated by the constraint) depend only on $\Omega_c$, which corresponds to the black hole mass in the Schwarzschild interior (see Ref. \cite{MM} for further details). 

On the other hand, for the first-order gauge invariant polar perturbations, we adopt a Fock representation, which is the conventional choice in quantum field theory in curved backgrounds. Our Hamiltonian formulation, which includes such gauge invariants as part of the phase space variables while isolating the purely gauge contributions, significantly simplifies the transition to the desired Fock description. Perturbative gauge constraints simply require that physical quantum states do not depend on the gauge sector of the perturbations. Then, focusing our attention on the gauge invariant sector, a more detailed analysis reveals that it is convenient to work with an alternative set of canonical invariant pairs. The reason is that, under reasonable conditions on the quantum dynamics of the perturbations, these new pairs guarantee the uniqueness of the Fock quantization \cite{AT, AT2}. 

\subsection{Uniqueness of the Fock quantization \label{sec: VA}}

Before addressing the Fock quantization of the gauge invariant polar modes, we want first to ensure that the Hamiltonian formulation is properly set up. To this end, we consider a canonical transformation that preserves the diagonal structure of the Hamiltonian for each gauge invariant mode (in perturbative configuration and momentum variables) while allowing us to select a specific time function for the associated equations of motion. For each mode, this time function is defined through the following expressions:
\begin{equation}
    k^2 = \omega_n^2 + l(l+1), \qquad \hat{l} = \frac{1}{k}\sqrt{l(l+1)}, \qquad k^2b_{\hat{l}}^2 = \omega_n^2p_c^2 + l(l+1)\frac{p_b^2}{L_o^2},
\end{equation}
where the parameter $k$ can be interpreted as a constant (angular) wave number in the space of mode labels $n$ and $l$, while $\hat{l}$ represents a normalized value relative to $k$. The subscript $\hat{l}$ in our notation indicates that the dependence on the mode labels is exclusively through $\hat{l}$. Based on these definitions, we present  the generating function of type-$2$ for the required canonical transformation,
\begin{equation}
    \mathscr{M}_{(2)}[\Tilde{Q}_i^{\mathfrak{n},\lambda},\mathscr{P}_i^{\mathfrak{n},\lambda}] = \sum_{\mathfrak{N}_2,\lambda}\bigg[\sum_{j=4}^{6}\Tilde{Q}_j^{\mathfrak{n},\lambda}\mathscr{P}_j^{\mathfrak{n},\lambda} + \sum_{j\in\{3,7\}}\bigg(\sqrt{b_{\hat{l}}}\Tilde{Q}_j^{\mathfrak{n},\lambda}\mathscr{P}_j^{\mathfrak{n},\lambda} - \frac{1}{4b_{\hat{l}}}b_{\hat{l}}'[\Tilde{Q}_j^{\mathfrak{n},\lambda}]^2\bigg)\bigg].
\end{equation}
The new canonical pairs are obtained from the relations
\begin{equation}
    \mathscr{Q}_i^{\mathfrak{n},\lambda} = \frac{\partial \mathscr{M}_{(2)}}{\partial \mathscr{P}_i^{\mathfrak{n},\lambda}}, \quad \Tilde{P}_i^{\mathfrak{n},\lambda} = \frac{\partial \mathscr{M}_{(2)}}{\partial \Tilde{Q}_i^{\mathfrak{n},\lambda}}.
\end{equation}
Since the canonical transformation does not alter the perturbative constraints, only the polar Hamiltonian is modified, following the standard formula for background-dependent canonical transformations, $\kappa\Tilde{\mathscr{H}}^{\text{po}} = \kappa\mathscr{H}^{\text{po}} + \mathscr{M}'$. Explicitly, its expression is
\begin{equation}
    \label{eq: VA-5.4}
    \Tilde{\mathscr{H}}^{\text{po}}[\Tilde{N}] = \sum_{\mathfrak{N}_2,\lambda} \tilde{N}\frac{b_{\hat{l}}}{2}\bigg[[\mathscr{P}_3^{\mathfrak{n},\lambda}]^2 + [k^2 + \mathfrak{u}][\mathscr{Q}_3^{\mathfrak{n},\lambda}]^2 + [\mathscr{P}_7^{\mathfrak{n},\lambda}]^2 + [k^2 + s_{\hat{l}}][\mathscr{Q}_7^{\mathfrak{n},\lambda}]^2\bigg].
\end{equation} 
This result closely resembles the perturbative Hamiltonian found in the FLRW model \cite{hyb1, hyb2, BGTT}. The background-dependent quantities $\mathfrak{u}$ and $s_{\hat{l}}$ act as mass terms for each perturbative mode and are given by
\begin{eqnarray}    
    \label{eq: VA-5.5}
     \begin{aligned}
         &\begin{aligned}
            s_{\hat{l}} = \frac{4}{b_{\hat{l}}^2}\left[\frac{p_b^2}{L_o^2} + \Omega_b^2\right] - 2\frac{\hat{l}^2}{b_{\hat{l}}^4}\frac{p_b^2}{L_o^2}\left[3(\Omega_b-\Omega_c)^2 + \frac{p_b^2}{L_o^2} + (\Omega_b^2-\Omega_c^2)\right] + 5\frac{\hat{l}^4}{b_{\hat{l}}^6}\frac{p_b^4}{L_o^4}(\Omega_b-\Omega_c)^2,
         \end{aligned}\\
         &\begin{aligned}
            \mathfrak{u} = s_{\hat{l}} + \frac{1}{b_{\hat{l}}^2}\bigg(\mathfrak{A}\mathfrak{B}-\frac{1}{4}\bigg[\mathfrak{C}-\frac{1}{2}\frac{\mathfrak{B}'}{\mathfrak{B}}\bigg]^2-\frac{1}{4}\bigg[\mathfrak{C}-\frac{1}{2}\frac{\mathfrak{B}'}{\mathfrak{B}}\bigg]' - k^2b_{\hat{l}}^2\bigg).
         \end{aligned}
     \end{aligned}
\end{eqnarray}

With a perturbative Hamiltonian of the form \eqref{eq: VA-5.4}, a privileged Fock quantization can be determined. For the pair $(\mathscr{Q}_7^{\mathfrak{n},\lambda},\mathscr{P}_7^{\mathfrak{n},\lambda})$, the result follows immediately from the discussion in Refs. \cite{AT,AT2} (since all the $k$-dependence in the Hamiltonian has been isolated into a single term). In more detail, it is ensured that the Fock quantization is unique (up to unitary equivalence) if we require invariance under the spatial isometries of the background and unitarity of the Heisenberg dynamics generated by the perturbative Hamiltonian when the background is treated as fixed. In contrast, analyzing the remaining pair requires a more careful approach, because isolating the wave number dependence in the Hamiltonian is not as straightforward as in the previous case. By expanding $\mathfrak{u}$ asymptotically in inverse powers of $k^2$, we find that $\mathfrak{u} = u_{\hat{l}} + O(k^{-2})$. Explicitly, the  zeroth-order contribution in the asymptotic expansion is given by
\begin{equation}
    u_{\hat{l}} = s_{\hat{l}} + \frac{4}{\hat{l}^2}\frac{L_o^2}{p_b^4}\bigg(2\Omega_c^2-13\Omega_b^2+10\Omega_b\Omega_c-4\frac{p_b^2}{L_o^2}\bigg) + \frac{4}{b_{\hat{l}}^2}\bigg(13\Omega_b^2-2\Omega_c^2+3\frac{p_b^2}{L_o^2}\bigg) + 4\frac{b_{\hat{l}}^2}{\hat{l}^4}\frac{L_o^4}{p_b^4}\bigg(2\Omega_b(\Omega_b-3\Omega_c)+\frac{p_b^2}{L_o^2}\bigg).
\end{equation}
With suitable modifications, the uniqueness proof of the Fock quantization presented in Ref. \cite{AT} can then be extended to this second gauge invariant pair under the same symmetry and unitarity requirements as before.

To introduce this Fock quantization, we construct creation and annihilationlike variables for each perturbative mode of the gauge invariant pairs through a mode-dependent linear combination of the configuration and momentum variables, with coefficients that depend only on the mode labels $n$ and $l$, but not on $m$. This last condition arises from the requirement of preserving the spatial isometries. Explicitly, we define the annihilationlike variables as
\begin{eqnarray}
    \label{eq: VA-5.7}
    a^{\mathfrak{n},\lambda}_{\text{G}} = f^{n,l}_{\text{G}}\mathscr{Q}_3^{\mathfrak{n},\lambda} + g^{n,l}_{\text{G}}\mathscr{P}_3^{\mathfrak{n},\lambda}, \qquad a^{\mathfrak{n},\lambda}_{\varphi} = f^{n,l}_{\varphi}\mathscr{Q}_7^{\mathfrak{n},\lambda} + g^{n,l}_{\varphi}\mathscr{P}_7^{\mathfrak{n},\lambda},
\end{eqnarray} 
where the subscript $\text{G}$ labels the modes associated with the geometric degrees of freedom of the perturbations, and similarly, the subscript $\varphi$ identifies the modes associated with the perturbative scalar field. The creationlike modes, namely $(a^{\mathfrak{n},\lambda}_{\text{G}})^*$ and $(a^{\mathfrak{n},\lambda}_{\varphi})^*$, are obtained by taking the complex conjugate (denoted by $*$) of the above expressions. To ensure that the transformed pairs fulfill the standard Poisson algebra of creation and annihilationlike variables, the coefficients must satisfy that 
\begin{eqnarray}
    \label{eq: VA-5.8}
    f^{n,l}_{\xi}(g^{n,l}_{\xi})^* - g^{n,l}_{\xi}(f^{n,l}_{\xi})^* = -i,
\end{eqnarray}
where we have introduced the compact notation $\xi= \text{G}$ or $\varphi$. Furthermore, to guarantee unitarity in the dynamics \cite{AT2}, the coefficients must behave as
\begin{eqnarray}
    \label{eq: VA-5.9}
    f^{n,l}_{\xi} = \sqrt{\frac{k}{2}} + k\vartheta^{n,l}_{(\xi,f)}, \qquad g^{n,l}_{\xi} = \frac{i}{\sqrt{2}k} + \vartheta^{n,l}_{(\xi,g)},
\end{eqnarray}  
where the $\vartheta$-functions represent subleading terms in $k$ that are constrained via
\begin{eqnarray}
    \label{eq: VA-5.10}
    \sum_{\mathfrak{N}_2}k\left|\vartheta^{n,l}_{(\xi,f)} + i\vartheta^{n,l}_{(\xi,g)}\right|^2 < \infty.
\end{eqnarray}  
In particular, this condition is satisfied by the so-called massless representation (namely, the natural representation in absence of the mass terms), which corresponds to vanishing $\vartheta$-functions in Eq. \eqref{eq: VA-5.9}. Therefore, we can choose this representation for the perturbations as a suitable representative of the family of equivalent Fock descriptions that satisfy our conditions of invariance under spatial symmetries and unitary Heisenberg evolution of the perturbations when the background is regarded as fixed.

\subsection{Quantum Hamiltonian constraint \label{sec: VB}} 

Once we have discussed the Fock quantization of the gauge invariant polar perturbations, we want to address the representation of the Hamiltonian constraint obtained from Eq. \eqref{eq: VA-5.4}. We recall that the representation of the background part of this constraint as an operator has already been accomplished within LQC (see e.g. Ref. \cite{MM}). The challenge to finding a quantum representation to the perturbative contribution \eqref{eq: VA-5.4} to this constraint is more intricate, because the coefficients of the corresponding quadratic perturbative terms are also background dependent. This background dependence arises through functions of the (densitized) triad variables and $\Omega$-variables (more specifically, through their positive and negative powers in both cases). In preparation for the quantization, it would be beneficial to eliminate quadratic or higher powers of $\Omega_b$ from the mass terms. Since $\Omega_c$ is a conserved quantity (and is represented by a Dirac observable in our LQC quantization), it is easier to handle at the quantum level, so its higher powers can be maintained. For any power of $\Omega_b$, we can reduce its degree by multiples of two using the background Hamiltonian, which vanishes up to quadratic contributions of the perturbations according to our Hamiltonian constraint, and therefore can be considered null in the perturbative contribution to this constraint. This strategy respects our perturbative truncation if it is followed by a suitable redefinition of the lapse function, similarly to what we have done in other parts of our discussion. Repeating this process as many times as necessary will leave only linear\footnote{If physical solutions are associated with a single root of the Hamiltonian among the two possible solutions for $\Omega_b$, even this linear contribution may be removed at the considered truncation order in perturbations. We will consider this possibility in future investigations.} or zero powers of $\Omega_b$. After this, the coefficients of the perturbative contribution to the Hamiltonian constraint can be expressed in terms of (essentially self-adjoint) operators defined in our loop quantization of the background. Since the $\Omega$-variables have a more complicated operator representation than the triad contributions, we must be cautious, specially when working with the logarithm of the strictly positive function $\mathfrak{B}$ to define the operator counterpart of $\mathfrak{B}'/\mathfrak{B}=\{\ln{\mathfrak{B}}, \Tilde{H}_{\text{KS}}\}_{\text{B}}$, with Poisson brackets translated into commutators [similar comments apply to $(\mathfrak{B}'/\mathfrak{B})'$]. We must also handle the ambiguity that arises when dealing with products of noncommuting operators within the same sector. To address this issue, we use a symmetric algebraic ordering for products of functions of $\Omega_b$ with nonnegative functions of $p_b$ (or $\Omega_c$ and $p_c$, respectively), multiplying the considered function of $\Omega_b$ (or $\Omega_c$) from both sides with the square root of the function of the triad. The outlined procedure ensures a consistent operator representation of the Hamiltonian within our hybrid-LQC approach.

Importantly, since the background in our analysis is dynamic, our approach accounts in principle for the backreaction of the perturbations. This happens because the Hamiltonian constraint is a global constraint at our truncation order on the entire system formed by the zero modes of the background and the gauge invariant (nonzero) modes of the perturbations. The construction of a consistently defined Hamiltonian constraint operator proves the nontriviality of the hybrid-LQC formalism. A final task, which will be the focus of future research, is to find quantum solutions to the Hamiltonian constraint (at least approximately) and derive master equations for the perturbations. For this, we propose to follow a method similar to that explained in Ref. \cite{LMM}, which introduces an ansatz with a suitable separation of variables and applies a mean field approximation to the background geometry.

\section{Vacuum state of the perturbations \label{sec: VI}}

In the previous section, we proved that the gauge invariant perturbations admit essentially a unique Fock quantization under the requirements that the spatial isometries of the background be respected and the Heisenberg evolution becomes unitary when the background can be treated as fixed, e.g. in effective regimes. Notice that the background effective trajectories need not be solutions in GR. However, we actually did not select any specific state in the Fock space of our quantization to play the role of the vacuum. In principle, any choice of creation and annihilationlike variables satisfying condition \eqref{eq: VA-5.10} is valid, and each of these choices picks out a concrete but different vacuum state, namely the state with unit norm on which all the corresponding annihilation operators have vanishing action. Such a vacuum state always belongs to the Fock space of our construction, because we have shown that all of our permitted choices of creation and annihilationlike variables determine unitarily equivalent representations. In this section, we introduce well-motivated criteria to fully restrict the freedom in that choice, so that a unique vacuum state is obtained for the gauge invariant polar perturbations.

In our discussion in Sec. \ref{sec: V}, we did not contemplate a possible dependence on the background of the $\vartheta$-functions that characterize the choice of creation and annihilationlike variables. The reason is that, for the analysis of the uniqueness of the Fock quantization (up to unitary equivalence), such dependence is not relevant. However, this issue becomes crucial for the choice of a vacuum state with good physical properties and dynamics. With this remark in mind, let us start our search by requiring that, for each of the two perturbative gauge invariant subsystems (the geometric sector and the scalar field sector), the action of the perturbative Hamiltonian operator, with the normal ordering corresponding to our choice of a vacuum, have a finite norm on such a state (treating the background as fixed). This finiteness is a reasonable requirement. Following arguments similar to those in Refs. \cite{AT2,BGTT}, it is not difficult to check that, if the coefficients in Eq. \eqref{eq: VA-5.7} were independent of the background, the resulting action of the Hamiltonian would be ill defined. To solve this problem, we allow that those coefficients and, as a result, the associated $\vartheta$-functions, depend on the background. In this new scenario, for each subsystem, the norm of the perturbative Hamiltonian operator acting on the vacuum remains finite, provided that the sum of the $\vartheta$-function associated with $f$ and $i$ times the $\vartheta$-function associated with $g$ is suitably proportional to the respective mass introduced in Eq. \eqref{eq: VA-5.5}, except for additional terms that behave asymptotically as $o(k^{-5/2})$. Here, the symbol $o(\cdot)$ denotes terms that are negligible with respect to its argument. For more details on this statement and its proof, the reader can consult Refs. \cite{AT2,BGTT}.

This result constrains the possible definitions of creation and annihilationlike variables, and hence of Fock representations. Additionally, by allowing the background dependence, we implicitly introduce a modification to the dynamics generated by the Hamiltonian. The reason is that, as in our previous discussion, the Hamiltonian is corrected owing to the dependence of the transformation on the background, which is not static. Although we do not present here the explicit expression of the corrected Hamiltonian, it is not difficult to derive it following the calculations presented in Ref. \cite{AT2}. 

In general, this new Hamiltonian contains an infinite set of self-interactions. Consequently, even though its normal-ordered action on the vacuum is well defined, it does not leave any domain of finite-particle states invariant. To further restrict the additional freedom introduced into our Fock description by considering the background-dependent transformations \eqref{eq: VA-5.7}, we propose imposing an asymptotic diagonalization of the perturbative Hamiltonian in the ultraviolet sector of infinitely large wave numbers $k$. That is, we require that our Hamiltonian, at least asymptotically, be proportional to the number operator. This requirement has already been implemented in other contexts, such as standard cosmology and perturbations around FLRW in LQC, leading to the elimination of the remaining freedom in the quantum system through the selection of a unique vacuum state \cite{BGTT,BGan}. Moreover, for backgrounds such as Minkowski or de Sitter spacetimes, one recovers in this way the natural vacuum for those situations \cite{BGTT}, namely the Poincar\'e and Bunch-Davies states, respectively, so that this requirement of asymptotic Hamiltonian diagonalization can be considered an extension of the most conventional criteria for the choice of a vacuum to spacetimes without symmetries involving the time direction.

The condition that the self-interaction terms of the Hamiltonian vanish in the asymptotic limit when $k$ takes large values allows us to fully constrain the coefficients of the linear transformation \eqref{eq: VA-5.7}. Firstly, following the discussion in Ref. \cite{AT2}, this condition fixes the relationship between the $\vartheta$-functions associated with $f$ and $g$. So, in the remainder, we only need to determine the solution for the $\vartheta$-function associated with $f$ and then substitute it into the aforementioned relationship to obtain the function for $g$. Additionally, using this relationship between the $\vartheta$-functions and Eq. \eqref{eq: VA-5.8}, which ensures the proper commutation relations for the creation and annihilationlike variables, we obtain a condition to determine both the modulus and the phase of the $\vartheta$-function associated with $f$, provided that we demand a minimum absorption of the background dynamics in this phase (see Ref. \cite{BGTT}). The explicit formulas are easy to obtain from a generalization of Ref. \cite{AT2}, and we do not provide them here to avoid unnecessary repetitions.

It is worth emphasizing that the background dependence in our definition of creation and annihilationlike variables can be expressed as an asymptotic series in $k$. Actually, if we define for each mode the quantity 
\begin{equation}
    \label{eq: VI-6.1}
    \chi_{\xi}^{n,l}= 2k^2 \left(1+i k \frac{g_{\xi}^{n,l}}{f_{\xi}^{n,l}}\right),
\end{equation}
and expand it as an asymptotic inverse-power series of the form
\begin{equation}
    \label{eq: VI-6.2}
    \chi_{\xi}^{n,l}= \sum_{j=0}^{\infty} \left(\frac{-i}{2 k}\right)^j \gamma_j^{\xi},
\end{equation}
an extension of the analysis carried out in Ref. \cite{AT2} shows that the background-dependent coefficients $\gamma_j^{\xi}$ can be determined recursively, starting from $j=0$, in terms of the mass term of the perturbations. For this, let us first express the mass term $\mathfrak{u}$ as the asymptotic series
\begin{equation}
    \mathfrak{u} = \sum_{j=0}^{\infty} \frac{u_{j}}{(2k)^j}.
\end{equation}
The other mass term, $s_{\hat{l}}$, is independent of $k$. The condition of asymptotic diagonalization of the Hamiltonian allows us to fix $\gamma_j^{\text{G}}$ by means of the recursive equation 
\begin{equation}
\gamma_j^{\text{G}} = -\frac{(\gamma_{j-1}^{\text{G}})'}{b_{\hat{l}}} + u_j - \sum_{p}^{j-2}\gamma_p^{\text{G}}\gamma_{j-p-2}^{\text{G}} + 4\sum_{p,q}^{p+q\leq j-4} \gamma_p^{\text{G}}\gamma_{j-p-q-4}^{\text{G}}u_q (-i)^{3q}+ 4\sum_{p}^{j-2}\gamma_{j-p-2}^{\text{G}}u_p (-i)^{3p}.
\end{equation}
A similar equation fixes $\gamma_j^{\varphi}$, replacing $u_j$ with $s_{\hat{l}}$ if $j=0$ and with zero if $j\geq 1$. The first coefficients in our series are $\gamma_0^{\text{G}}=u_{\hat{l}}$ and $\gamma_0^{\varphi}=s_{\hat{l}}$. We recall that the prime denotes the Poisson bracket with the background contribution to the Hamiltonian constraint. This procedure completely determines the choice of canonical variables for the Fock quantization, at least asymptotically. Physically, this choice corresponds to the selection of a unique vacuum state. The corresponding vacuum solution can be constructed using Eqs. \eqref{eq: VI-6.1} and \eqref{eq: VI-6.2} and following the same steps as in Ref. \cite{AT2}.

\section{Conclusions \label{sec: VII}}

With this work, we can conclude that the study of perturbative modes in the Kantowski-Sachs spacetime, at least up to first order, admits a unified Hamiltonian formulation. Until now, this formulation was only applicable to a specific sector of the model, corresponding only to axial perturbations, and it was unclear whether it could be extended to the entire system, including also polar perturbations.  

Our results confirm that a Hamiltonian analysis of the perturbed Kantowski-Sachs spacetime is indeed feasible through the study of its truncated action. By retaining terms up to quadratic perturbative order, we have derived an action composed of two distinct contributions: one corresponding to the background (zeroth order) and another accounting for the linear perturbations (pure quadratic order). Notably, the term at linear order in the expansion ultimately vanishes. For the background treatment, we have adopted variables inspired by LQG. While this choice is not essential for developing the classical formulation, it facilitates the connection between our results and the proposed quantum analysis. The study of linear perturbations arises from an effective quadratic term in the action, which exclusively receives contributions from the product of two linear perturbations. The analysis of perturbations is significantly streamlined by expanding them in real Fourier and spherical harmonics. This decomposition has allowed us to isolate the truly relevant physical information within the real coefficients of the expansion.

Applying Hamiltonian techniques to the truncated action, we can observe that the symmetries of the model emerge through the presence of constraints. These constraints, which involve both the background and the perturbations, restrict the allowed values of the phase space variables and ensure that only gauge invariant perturbations correspond to physical perturbative degrees of freedom. Mathematically, these first-order perturbative gauge invariants are defined as perturbative quantities that remain invariant under the Hamiltonian flow generated by the perturbative constraints of the system. The choice of phase space variables can be made such that the background and the perturbative degrees of freedom form a global canonical set, a key property that plays a fundamental role in the transition to the quantum analysis. Within this phase space, the perturbative sector that we have studied is known as the polar sector. It is characterized by five canonical pairs and three constraints per perturbative mode, implying that only two of the five pairs contribute to physically relevant results.

By following the ideas outlined in Refs. \cite{M,DC}, we have isolated the perturbative physical degrees of freedom through a series of well-chosen canonical transformations. The key idea is to change from the initial set to a new one that consists of two gauge invariant pairs, alongside three pairs formed by the perturbative constraints and their associated momenta. These canonical transformations affect each perturbative mode individually and depend on the background variables. Less directly, they also affect the background variables themselves, as maintaining the global canonical structure (including both background and perturbations) requires corrections of the background variables involving quadratic terms in the perturbations. While these redefinitions are necessary, they do not alter the results at the perturbative order we are working with, reaching the same conclusions if we simply keep in mind that the uncorrected background variables are in practice replaced with corrected ones.

For the first time, and without fixing the gauge at any stage, we have determined the canonical pairs that encode the true physical information of the Kantowski-Sachs polar perturbations. In the Hamiltonian formulation, it becomes clear which pairs carry this information and which correspond to pure gauge degrees of freedom. Although the changes of phase space variables may seem to complicate the dynamics, the freedom to redefine the Lagrange multipliers associated with the perturbative constraints and the Hamiltonian constraint allows us to simplify the description. By introducing perturbative corrections to these multipliers, we can eliminate any gauge dependent contributions from the perturbative Hamiltonian at our order of truncation. The resulting gauge invariant Hamiltonian, which we refer to as the true (polar) perturbative Hamiltonian, has a particularly simple structure, resembling a double sequence of harmonic oscillators with background-dependent frequencies. This significantly simplifies the computation of the dynamics of the gauge invariant polar pairs and provides a strong motivation for the quantum study of the model.

We have proposed the hybrid formalism as a suitable framework for the quantization of our Kantowski-Sachs model with perturbations. Guided by the remarkable properties that the loop quantization offers for treating the background degrees of freedom, we have taken into account the results presented in Ref. \cite{GBA} and combined them with a Fock quantization of the perturbative gauge invariants. The viability of a privileged Fock representation (up to unitary equivalence) is supported by the results of Refs. \cite{AT, AT2, BGTT}. Although these works analyze cosmological scenarios different from ours, their conclusions can be extended through a convenient generalization. Unlike the axial case (see Ref. \cite{MM}), the application of those works to our model is not as straightforward, because one of the mass terms for the polar modes is more challenging to handle. To solve these complications, we have conducted a careful asymptotic analysis of the term in question. In this way, we have confirmed that, if the background isometries are preserved and we enforce unitary Heisenberg dynamics when the background is regarded as fixed, the Fock representation for the polar modes is unique up to unitarity. Additionally, by imposing further restrictions, we can refine this family of representations and select a privileged vacuum state, at least in regimes where the background can be treated effectively or in a mean field approximation. This choice of vacuum guarantees that there is no Hamiltonian particle production in the ultraviolet.

We have also argued that the quantum representation of the perturbative constraints and the Hamiltonian constraint is within reach. On one hand, the action of the perturbative constraints implies that the physical states are independent of the purely gauge modes of the perturbations. On the other hand, despite the potential complexity of the Hamiltonian, owing to intricate background-dependent coefficients in its perturbative terms, it can be managed using the LQC techniques developed in Ref. \cite{MM} and commented in this work. These positive aspects make the hybrid formalism a strong candidate for continuing the quantum study of the perturbations.

A more innovative avenue would involve identifying our first-order perturbative gauge invariants, particularly those relevant for quantization, in terms of more standard invariants used in the literature of Schwarzschild-type black holes. This approach could extend our results beyond the black hole interior, potentially offering new insights into gravitational radiation, and the final stages of black hole formation. The similarities between the axial and polar results further motivate us to refine our classical formulation and seek transformations that relate both sectors, following a line of research similar to that of Refs. \cite{CM1, CM2}. We also expect that these results could naturally lead to an exact analysis of the complicated mass term in the polar sector. Given its relative simplicity, starting with the axial sector to carry out the first steps of the analysis appears to be the most natural approach. Finally, future work should explore the hybrid formalism more deeply, aiming to compute quantum corrections to observable predictions, following the approach taken in cosmological studies \cite{hyb-review,hyb1,AAN2}.

\acknowledgments

The authors are grateful to B. Elizaga Navascu\'es and A. Torres-Caballeros for discussions. This work was supported by the Spanish Grant No. PID2020-118159GB-C41 funded by MCIN/AEI/10.13039/501100011033/ and partially by Project No. 2024AEP005 from the Spanish National Research Council (CSIC). A. M.-S. acknowledges support from the PIPF-2023 fellowship from Comunidad Aut\'onoma de Madrid. The reference number is PIPF-2023/TEC-30167.

\newpage

\appendix

\section{CALCULATIONS FOR THE POLAR PERTURBATIONS \label{sec: app-1}}

This appendix contains the expressions of the first-order perturbative gauge constraints, presented in Eq. \eqref{eq: III-3.5}. They are given by
\begin{equation}
    \label{eq: app-A1}
    \begin{aligned}
        &\begin{aligned}
            \mathbf{C}_1[q_0^{\mathfrak{n},\lambda}] &= \sum_{\mathfrak{N}_1,\lambda} q_0^{\mathfrak{n},\lambda}l(l+1)\bigg[ - 2\frac{L_o^2}{p_b^2}\Omega_b|p_c|h_6^{\mathfrak{n},\lambda} - |p_c|p_3^{\mathfrak{n},\lambda} + \lambda w_n |p_c|\bigg(4\frac{L_o^2}{p_b^2}\Omega_bh_5^{\mathfrak{n},-\lambda} - \frac{p_5^{\mathfrak{n},-\lambda}}{l(l+1)}\bigg)\bigg]\\
            &+ \sum_{\mathfrak{N}_2,\lambda} q_0^{\mathfrak{n},\lambda}\bigg[2|p_c|p_4^{\mathfrak{n},\lambda} - \frac{(l+2)!}{(l-2)!}(\Omega_b+\Omega_c)\frac{h_4^{\mathfrak{n},\lambda}}{|p_c|}\bigg],
        \end{aligned}\\
        &\begin{aligned}
            \mathbf{C}_2[k_0^{\mathfrak{n},\lambda}] &= \sum_{\mathfrak{N}_0,\lambda} k_0^{\mathfrak{n},\lambda}2\lambda\omega_n|p_c|\bigg[\frac{L_o^2}{p_b^2}\Omega_b|p_c|h_6^{\mathfrak{n},-\lambda} - (\Omega_b+\Omega_c)\frac{h_3^{\mathfrak{n},-\lambda}}{|p_c|} - \frac{p_b^2}{L_o^2}\frac{p_6^{\mathfrak{n},-\lambda}}{|p_c|}\bigg]\\
            &+ \sum_{\mathfrak{N}_1,\lambda} k_0^{\mathfrak{n},\lambda}\bigg[\frac{p_b^2}{L_o^2}p_5^{\mathfrak{n},\lambda} - 2l(l+1)(\Omega_b+\Omega_c)h_5^{\mathfrak{n},\lambda}\bigg],
        \end{aligned}\\
&\begin{aligned}
            C_3[f_0^{\mathfrak{n},\lambda}] &= -\sum_{\mathfrak{N}_0,\lambda} \Tilde{N}f_0^{\mathfrak{n},\lambda}\bigg[\Omega_b|p_c|p_3^{\mathfrak{n},\lambda} - \bigg(\frac{l^2+l+2}{2}\frac{p_b^2}{L_o^2} + \omega_n^2p_c^2 + 2\Omega_b^2 + 2\Omega_b\Omega_c + 2\frac{\Tilde{H}_{\text{KS}}}{L_o}\bigg)\frac{h_3^{\mathfrak{n},\lambda}}{|p_c|} + \frac{p_b^2}{L_o^2}\Omega_c\frac{p_6^{\mathfrak{n},\lambda}}{|p_c|}\\
            &- \frac{L_o^2}{p_b^2}\bigg(2\Omega_b\Omega_c + \frac{l^2+l+2}{2}\frac{p_b^2}{L_o^2} + \frac{\Tilde{H}_{\text{KS}}}{L_o}\bigg)|p_c|h_6^{\mathfrak{n},\lambda}\bigg] - \sum_{\mathfrak{N}_1,\lambda} \Tilde{N}f_0^{\mathfrak{n},\lambda}\lambda\omega_n|p_c|l(l+1)h_5^{\mathfrak{n},-\lambda}\\
            &+ \sum_{\mathfrak{N}_2,\lambda} \frac{\Tilde{N}}{4}f_0^{\mathfrak{n},\lambda}\frac{(l+2)!}{(l-2)!}\frac{p_b^2}{L_o^2}\frac{h_4^{\mathfrak{n},\lambda}}{|p_c|}.
        \end{aligned}
\end{aligned}
\end{equation}
In addition, the value of the polar perturbative Hamiltonian, after spatial integration, is
\begin{equation}
    \begin{aligned}
        \kappa \Tilde{H}^{\text{po}}[\Tilde{N}] &= \sum_{\mathfrak{N}_0,\lambda}\frac{\Tilde{N}}{2}\bigg[ [p_7^{\mathfrak{n},\lambda}]^2 + \bigg(\omega_n^2p_c^2+l(l+1)\frac{p_b^2}{L_o^2}\bigg)[h_7^{\mathfrak{n},\lambda}]^2 + \bigg(8\Omega_b(\Omega_b+\Omega_c)+4\frac{p_b^2}{L_o^2}-\omega_n^2p_c^2+8\frac{\Tilde{H}_{\text{KS}}}{L_o}\bigg)\frac{[h_3^{\mathfrak{n},\lambda}]^2}{p_c^2}\\
        &+ \frac{p_b^4}{L_o^4}\frac{[p_6^{\mathfrak{n},\lambda}]^2}{p_c^2} + \bigg(4\Omega_b(\Omega_b+\Omega_c)+2\frac{p_b^2}{L_o^2}+3\frac{\Tilde{H}_{\text{KS}}}{L_o}\bigg)\frac{L_o^4}{p_b^4}p_c^2[h_6^{\mathfrak{n},\lambda}]^2 + 2\frac{p_b^2}{L_o^2}\bigg(\frac{2}{p_c^2}\Omega_bh_3^{\mathfrak{n},\lambda} - p_3^{\mathfrak{n},\lambda}\bigg)p_6^{\mathfrak{n},\lambda}- \frac{L_o^2}{p_b^2}\\
        &\times\bigg(4\Omega_b(\Omega_b-\Omega_c)+(l+2)(l-1)\frac{p_b^2}{L_o^2}-4\frac{\Tilde{H}_{\text{KS}}}{L_o}\bigg)h_3^{\mathfrak{n},\lambda}h_6^{\mathfrak{n},\lambda} - (4\Omega_b-2\Omega_c)h_6^{\mathfrak{n},\lambda}p_6^{\mathfrak{n},\lambda} + 2\frac{L_o^2}{p_b^2}p_c^2\Omega_bh_6^{\mathfrak{n},\lambda}p_3^{\mathfrak{n},\lambda} \bigg] \\
        & + \sum_{\mathfrak{N}_1,\lambda}\frac{\Tilde{N}}{2}\bigg[2\lambda\omega_nl(l+1)h_3^{\mathfrak{n},\lambda}h_5^{\mathfrak{n},-\lambda} + \frac{p_b^2}{L_o^2}\frac{[p_5^{\mathfrak{n},\lambda}]^2}{l(l+1)}+ 2\frac{L_o^2}{p_b^2}l(l+1)\bigg(4\Omega_b(\Omega_b+\Omega_c)+\frac{p_b^2}{L_o^2}+2\frac{\Tilde{H}_{\text{KS}}}{L_o}\bigg)[h_5^{\mathfrak{n},\lambda}]^2 \\
        & - 4\Omega_bh_5^{\mathfrak{n},\lambda}p_5^{\mathfrak{n},\lambda}\bigg] + \sum_{\mathfrak{N}_2,\lambda} \frac{\Tilde{N}}{2}\bigg[4p_c^2\frac{(l-2)!}{(l+2)!}[p_4^{\mathfrak{n},\lambda}]^2+ \frac{1}{4p_c^2}\frac{(l+2)!}{(l-2)!}\bigg(4(\Omega_b+\Omega_c)^2+2\frac{p_b^2}{L_o^2}+\omega_n^2p_c^2+4\frac{\Tilde{H}_{\text{KS}}}{L_o}\bigg)\\
        &\times[h_4^{\mathfrak{n},\lambda}]^2 - 4\Omega_ch_4^{\mathfrak{n},\lambda}p_4^{\mathfrak{n},\lambda} - \frac{1}{2}\frac{(l+2)!}{(l-2)!}h_6^{\mathfrak{n},\lambda}h_4^{\mathfrak{n},\lambda}
        - \frac{(l+2)!}{(l-2)!}\lambda\omega_nh_5^{\mathfrak{n},\lambda}h_4^{\mathfrak{n},-\lambda}\bigg].
    \end{aligned}
\end{equation}
We have used the same notation as in the main text. We recall that $\lambda\in\{+,-\}\simeq\mathbb{Z}_2$, so that, when we write $-\lambda$, we refer to the opposite element of $\lambda$, regardless of its value.

\section{CALCULATION FOR THE PERTURBATIVE GAUGE INVARIANTS \label{sec: app-2}}

This appendix completes the results presented in Sec. \ref{sec: IV}. Concerning the canonical transformation \eqref{eq: IV-4.1}, we have the following redefinitions. The third perturbative constraint can be expressed as
\begin{equation}
    \label{eq: app-B.1}
    \Tilde{\mathbf{C}}_3[f_0^{\mathfrak{n},\lambda}] = \mathbf{C}_3[f_0^{\mathfrak{n},\lambda}] - \sum_{\mathfrak{N}_2,\lambda} 2\Tilde{N}f_0^{\mathfrak{n},\lambda}\big( \Bar{h}_4^{\mathfrak{n},\lambda} + 2\Bar{h}_3^{\mathfrak{n},\lambda}\big)\frac{\Tilde{H}_{\text{KS}}}{L_o}.
\end{equation}
The expression of the redefined lapse function consists of two contributions, namely, one that cancels the first block of $K^{\text{po}}$ (proportional to the background Hamiltonian) and another that compensates for the redefinition given in Eq. \eqref{eq: app-B.1}. Explicitly, we have
\begin{equation}
    \begin{aligned}
        \bar{N} &= \hat{N} + \frac{\epsilon^2}{\kappa L_o}\sum_{\mathfrak{N}_2,\lambda}\bigg[2\lambda\omega_n|p_c|(\bar{h}_4^{\mathfrak{n},\lambda} - 2\bar{h}_3^{\mathfrak{n},\lambda})\bar{h}_6^{\mathfrak{n},-\lambda} + \frac{l(l+1)}{2}\frac{L_o^2}{p_b^2}\bigg(\omega_n^2p_c^2+(l+2)(l-1)\frac{p_b^2}{L_o^2}\bigg)[\bar{h}_5^{\mathfrak{n},\lambda}]^2\\
        &+ l(l+1)\bar{h}_4^{\mathfrak{n},\lambda}\bar{h}_5^{\mathfrak{n},\lambda} + 2\bigg(\omega_n^2p_c^2+l(l+1)\frac{p_b^2}{L_o^2}\bigg)[\bar{h}_6^{\mathfrak{n},\lambda}]^2\bigg] + \epsilon^2\sum_{\mathfrak{N}_2,\lambda} 2\Tilde{N}\frac{f_0^{\mathfrak{n},\lambda}}{L_o}\big(\Bar{h}_4^{\mathfrak{n},\lambda} + 2\Bar{h}_3^{\mathfrak{n},\lambda}\big).
    \end{aligned}
\end{equation}
The perturbative Lagrange multipliers for the first two constraints, after eliminating the second block of the corrected perturbative Hamiltonian (proportional to those constraints), are defined as follows:
\begin{equation}
    \begin{aligned}
        &\begin{aligned}
            \kappa\mathbf{q}_0^{\mathfrak{n},\lambda} &= \kappa q_0^{\mathfrak{n},\lambda} + 2\Tilde{N}\frac{(l-2)!}{(l+2)!}\frac{L_o^2}{p_b^2}\bigg[\omega_n^2p_c^2\left(\Omega_b\bar{h}_4^{\mathfrak{n},\lambda}-(\Omega_b+\Omega_c)\bar{h}_3^{\mathfrak{n},\lambda} - \bar{p}_4^{\mathfrak{n},\lambda}\right)\\
            &+ l(l+1)\frac{p_b^2}{L_o^2}\bigg(\Omega_b\bar{h}_4^{\mathfrak{n},\lambda}+\frac{1}{4}\bar{p}_3^{\mathfrak{n},\lambda} + \frac{1}{2}\frac{\bar{p}_5^{\mathfrak{n},\lambda}}{l(l+1)}\bigg) + \frac{1}{2}\lambda\omega_n|p_c|\bar{p}_6^{\mathfrak{n},-\lambda}\bigg],
        \end{aligned}\\
        &\begin{aligned}
            \kappa\mathbf{k}_0^{\mathfrak{n},\lambda} &= \kappa k_0^{\mathfrak{n},\lambda} - \frac{\Tilde{N}}{2}\bigg[4\lambda\omega_n|p_c|\frac{(l-2)!}{(l+2)!}\frac{L_o^4}{p_b^4}\bigg(\bigg[\omega_n^2p_c^2+(l+2)(l-1)\frac{p_b^2}{L_o^2}\bigg]\left[\Omega_b\bar{h}_4^{\mathfrak{n},-\lambda} - (\Omega_b+\Omega_c)\bar{h}_3^{\mathfrak{n},-\lambda}\right]\\
            &+ \frac{1}{2}l(l+1)\frac{p_b^2}{L_o^2}\left[2\Omega_b\bar{h}_4^{\mathfrak{n},-\lambda} + \bar{p}_3^{\mathfrak{n},-\lambda}\right] - \bigg[\omega_n^2p_c^2+\frac{1}{2}(l+2)(l-1)\frac{p_b^2}{L_o^2}\bigg]\bar{p}_4^{\mathfrak{n},-\lambda}\bigg) + 4\Omega_b\bar{h}_6^{\mathfrak{n},\lambda}\\
            &- \frac{(l-2)!}{(l+2)!}\frac{L_o^4}{p_b^4}\bigg(\omega_n^2p_c^2+(l+2)(l-1)\frac{p_b^2}{L_o^2}\bigg)\bar{p}_6^{\mathfrak{n},\lambda}\bigg].
        \end{aligned}
    \end{aligned}
\end{equation}
Finally, the expression of the third-block term, which we also refer to as the true polar perturbative Hamiltonian, is
\begin{equation}
    \begin{aligned}
        \kappa \bar{H}^{\text{po}}[\Tilde{N}] &= \sum_{\mathfrak{N}_2,\lambda} \frac{\Tilde{N}}{2}\bigg[ \bigg(1+4\omega_n^2p_c^2\frac{(l-2)!}{(l+2)!}\frac{L_o^4}{p_b^4}\bigg[\omega_n^2p_c^2+(l+2)(l-1)\frac{p_b^2}{L_o^2}\bigg]\bigg)[\bar{p}_4^{\mathfrak{n},\lambda}]^2 + \frac{l(l+1)}{(l+2)(l-1)}[\bar{p}_3^{\mathfrak{n},\lambda}]^2\\
        &- \bigg(2\Omega_c+4\Omega_b\bigg[1 + 2\omega_n^2p_c^2\frac{(l-2)!}{(l+2)!}\frac{L_o^4}{p_b^4}\bigg(\omega_n^2p_c^2+(l+2)(l-1)\frac{p_b^2}{L_o^2}+l(l+1)\frac{p_b^2}{L_o^2}\bigg)\bigg]\bigg)\bar{p}_4^{\mathfrak{n},\lambda}\bar{h}_4^{\mathfrak{n},\lambda}\\
        &+\bigg(4\Omega_b+8\omega_n^2p_c^2\frac{(l-2)!}{(l+2)!}\frac{L_o^4}{p_b^4}(\Omega_b+\Omega_c)\bigg[\omega_n^2p_c^2+(l+2)(l-1)\frac{p_b^2}{L_o^2}\bigg]\bigg)\bar{p}_4^{\mathfrak{n},\lambda}\bar{h}_3^{\mathfrak{n},\lambda} + \bigg(4\Omega_b(\Omega_b+\Omega_c)\\
        &+ 2\frac{p_b^2}{L_o^2} + 3\frac{\Tilde{H}_{\text{KS}}}{L_o} + 4\frac{(l-2)!}{(l+2)!}\frac{L_o^4}{p_b^4}\Omega_b^2\bigg[\bigg(\omega_n^2p_c^2+l(l+1)\frac{p_b^2}{L_o^2}\bigg)^2 + (l+2)(l-1)\frac{p_b^2}{L_o^2}\omega_n^2p_c^2\bigg]\bigg)[\bar{h}_4^{\mathfrak{n},\lambda}]^2\\
        &-\bigg(4\Omega_b+4\omega_n^2p_c^2\frac{(l-2)!}{(l+2)!}l(l+1)\frac{L_o^2}{p_b^2}(\Omega_b+\Omega_c)\bigg)\bar{p}_3^{\mathfrak{n},\lambda}\bar{h}_3^{\mathfrak{n},\lambda} - \bigg(4\Omega_b(\Omega_b-\Omega_c)+(l+2)(l-1)\frac{p_b^2}{L_o^2}\\
        &- 4\frac{\Tilde{H}_{\text{KS}}}{L_o} + 8\omega_n^2p_c^2\frac{(l-2)!}{(l+2)!}\frac{L_o^4}{p_b^4}\Omega_b(\Omega_b+\Omega_c)\bigg[\omega_n^2p_c^2+(l+2)(l-1)\frac{p_b^2}{L_o^2}+l(l+1)\frac{p_b^2}{L_o^2}\bigg]\bigg)\bar{h}_4^{\mathfrak{n},\lambda}\bar{h}_3^{\mathfrak{n},\lambda}\\
        & - 2\bigg(1+2\omega_n^2p_c^2\frac{(l-2)!}{(l+2)!}l(l+1)\frac{L_o^2}{p_b^2}\bigg)\bar{p}_4^{\mathfrak{n},\lambda}\bar{p}_3^{\mathfrak{n},\lambda} + \bigg(8\Omega_b(\Omega_b+\Omega_c)+4\frac{p_b^2}{L_o^2}-\omega_n^2p_c^2+8\frac{\Tilde{H}_{\text{KS}}}{L_o}\\
        &+ 4\omega_n^2p_c^2\frac{(l-2)!}{(l+2)!}\frac{L_o^4}{p_b^4}(\Omega_b+\Omega_c)^2 \bigg[\omega_n^2p_c^2+(l+2)(l-1)\frac{p_b^2}{L_o^2}\bigg]\bigg)[\bar{h}_3^{\mathfrak{n},\lambda}]^2 + \bigg(\omega_n^2p_c^2+l(l+1)\frac{p_b^2}{L_o^2}\bigg)[\bar{h}_7^{\mathfrak{n},\lambda}]^2\\
        &+ [\bar{p}_7^{\mathfrak{n},\lambda}]^2 + 2\Omega_b\bigg(1+2\frac{(l-2)!}{(l+2)!}l(l+1)\frac{L_o^2}{p_b^2}\bigg[\omega_n^2p_c^2+l(l+1)\frac{p_b^2}{L_o^2}\bigg]\bigg)\bar{h}_4^{\mathfrak{n},\lambda}\bar{p}_3^{\mathfrak{n},\lambda}\bigg].
    \end{aligned}
\end{equation}

Regarding the additional canonical transformation given by Eq. \eqref{eq: IV-4.6}, we need redefinitions that follow a criterion similar to that explained above for the previous transformation. In this sense, the new lapse function, after eliminating the block of $\textbf{H}^{\text{po}}$ proportional to the background Hamiltonian, is defined as follows:
\begin{equation}
    \begin{aligned}
        \mathbf{\Tilde{N}} &= \bar{N} + \frac{\epsilon^2}{\kappa L_o}\sum_{\mathfrak{N}_2,\lambda}\bigg[\bigg(8\frac{(l-2)!}{(l+2)!}\frac{L_o^4}{p_b^4}\bigg[\bigg(\omega_n^2-l(l+1)\frac{p_b^2}{L_o^2}\bigg)^2+(l+2)(l-1)\frac{p_b^2}{L_o^2}\omega_n^2\bigg]\bigg[ \bigg(\bigg[\Omega_b\omega_n^2p_c^2 - \Omega_c\frac{l^2+l+2}{2}\frac{p_b^2}{L_o^2}\bigg]\\
        &\times Q_3^{\mathfrak{n},\lambda} - \Omega_b^2P_3^{\mathfrak{n},\lambda}\bigg) + \frac{\Tilde{H}_{\text{KS}}}{2L_o}Q_4^{\mathfrak{n},\lambda}\bigg] + \bigg[4\Omega_b^2+2\Omega_b\Omega_c+\frac{3}{2}\Omega_c^2-\omega_n^2p_c^2-l(l+1)\frac{p_b^2}{L_o^2}\bigg]Q_4^{\mathfrak{n},\lambda} + \bigg[\omega_n^2p_c^2 - l(l+1)\frac{p_b^2}{L_o^2}\bigg]\\
        &\times4\bigg[(l+2)(l-1)\frac{p_b^2}{L_o^2}\bigg]^{-1}\bigg[\bigg(\frac{l^2+l+2}{2}\frac{p_b^2}{L_o^2} + \omega_n^2p_c^2\bigg)Q_4^{\mathfrak{n},\lambda} + P_3^{\mathfrak{n},\lambda}\bigg] + 2\Omega_b(4\Omega_b+\Omega_c)\bigg[\frac{l^2+l+2}{2}\frac{p_b^2}{L_o^2} + \omega_n^2p_c^2\bigg]^{-1}\\
        &\times P_3^{\mathfrak{n},\lambda} + \bigg[\Omega_bP_3^{\mathfrak{n},\lambda} - \bigg(\frac{l^2+l+2}{2}\frac{p_b^2}{L_o^2} + \omega_n^2p_c^2\bigg)Q_3^{\mathfrak{n},\lambda}\bigg]\bigg[(8\Omega_b-2\Omega_c)\bigg(\frac{l^2+l+2}{2}\frac{p_b^2}{L_o^2} + \omega_n^2p_c^2\bigg)^{-1} + (2\Omega_b+3\Omega_c)\\
        &\times\bigg(\frac{l^2+l+2}{2}\frac{p_b^2}{L_o^2}\bigg)^{-1}\bigg]\bigg)Q_4^{\mathfrak{n},\lambda} + \frac{3}{2}\bigg(\frac{l^2+l+2}{2}\frac{p_b^2}{L_o^2}\bigg)^{-2}\bigg(\Omega_bP_3^{\mathfrak{n},\lambda} - \bigg[\frac{l^2+l+2}{2}\frac{p_b^2}{L_o^2} + \omega_n^2p_c^2\bigg]Q_3^{\mathfrak{n},\lambda}\bigg)^2 + 2\Omega_b\\
        &\times\bigg(\frac{l^2+l+2}{2}\frac{p_b^2}{L_o^2}\bigg)^{-1}Q_3^{\mathfrak{n},\lambda}P_3^{\mathfrak{n},\lambda} + 2\bigg(1-\omega_n^2p_c^2\bigg[\frac{l^2+l+2}{2}\frac{p_b^2}{L_o^2}\bigg]^{-1}\bigg)[Q_3^{\mathfrak{n},\lambda}]^2\bigg].
    \end{aligned}
\end{equation}
The redefined perturbative Lagrange multiplier associated with the third constraint is given by
\begin{equation}
    \begin{aligned}
        2\kappa \mathbf{f}_0^{\mathfrak{n},\lambda} &= 2\kappa f_0^{\mathfrak{n},\lambda} - 4\frac{(l-2)!}{(l+2)!}\frac{L_o^4}{p_b^4}\bigg[\bigg(\omega_n^2-l(l+1)\frac{p_b^2}{L_o^2}\bigg)^2+(l+2)(l-1)\frac{p_b^2}{L_o^2}\omega_n^2\bigg]2\Omega_b\bigg[(l^2+l+2)\frac{p_b^2}{L_o^2}\frac{\Tilde{H}_{\text{KS}}}{L_o}Q_4^{\mathfrak{n},\lambda}\\
        &+ \bigg(\bigg[\Omega_b\omega_n^2p_c^2 - \Omega_c\frac{l^2+l+2}{2}\frac{p_b^2}{L_o^2}\bigg] Q_3^{\mathfrak{n},\lambda} - \Omega_b^2P_3^{\mathfrak{n},\lambda}\bigg) + \frac{\Omega_b}{2} P_4^{\mathfrak{n},\lambda}\bigg]\bigg(\frac{l^2+l+2}{2}\frac{p_b^2}{L_o^2}\bigg)^{-2} - 2(2\Omega_b+\Omega_c)Q_4^{\mathfrak{n},\lambda}\\
        &+ 4\Omega_b\bigg(\omega_n^2p_c^2 - l(l+1)\frac{p_b^2}{L_o^2}\bigg)\bigg((l+2)(l-1)\frac{p_b^2}{L_o^2}\bigg)^{-1}\bigg(\frac{l^2+l+2}{2}\frac{p_b^2}{L_o^2}\bigg)^{-1}P_3^{\mathfrak{n},\lambda} + \bigg(\frac{l^2+l+2}{2}\frac{p_b^2}{L_o^2}\bigg)^{-1}\\
        &\times\bigg[ 2\Omega_bP_3^{\mathfrak{n},\lambda} + (l+2)(l-1)\frac{p_b^2}{L_o^2}Q_3^{\mathfrak{n},\lambda} - \bigg(6\bigg[\bigg(\Omega_bP_3^{\mathfrak{n},\lambda} - \bigg[\frac{l^2+l+2}{2}\frac{p_b^2}{L_o^2} + \omega_n^2p_c^2\bigg]Q_3^{\mathfrak{n},\lambda}\bigg)-\frac{1}{2}P_4^{\mathfrak{n},\lambda}\bigg]\\
        &\times \bigg[\frac{l^2+l+2}{2}\frac{p_b^2}{L_o^2}\bigg]^{-1} + (4\Omega_b+6\Omega_c)Q_4^{\mathfrak{n},\lambda} + 4Q_3^{\mathfrak{n},\lambda} \bigg)\frac{\Tilde{H}_{\text{KS}}}{L_o}\bigg]
    \end{aligned}
\end{equation}
Finally, the three background-dependent and mode-dependent expressions appearing in Eq. \eqref{eq: IV-4.9} are
\begin{equation}
    \begin{aligned}
        &\begin{aligned}
            \mathfrak{A} &= \bigg(\omega_n^2p_c^2 + l(l+1)\frac{p_b^2}{L_o^2}\bigg) - \frac{8\omega_n^2p_c^2}{l^2+l+2} + 4\frac{(l-2)!}{(l+2)!}\frac{L_o^4}{p_b^4}\bigg(\Omega_b\omega_n^2p_c^2 - \Omega_c\frac{l^2+l+2}{2}\frac{p_b^2}{L_o^2}\bigg)^2\\
            &\times\bigg[\bigg(\omega_n^2p_c^2 - l(l+1)\frac{p_b^2}{L_o^2}\bigg)^2 + (l+2)(l-1)\frac{p_b^2}{L_o^2}\omega_n^2p_c^2\bigg]\bigg(\frac{l^2+l+2}{2}\frac{p_b^2}{L_o^2}\bigg)^{-2},
        \end{aligned}\\
        &\begin{aligned}
            \mathfrak{B} &= \frac{l(l+1)}{(l+2)(l-2)}\bigg[1-\frac{2\Omega_b^2}{l(l+1)}\bigg(\omega_n^2p_c^2 - \frac{l^2+l+2}{2}\frac{p_b^2}{L_o^2}\bigg)\bigg(\frac{l^2+l+2}{2}\frac{p_b^4}{L_o^4}\bigg)^{-1}\bigg]^2\\
            &+\Omega_b^4\frac{3l^2+3l+2}{l(l+1)}\bigg(\frac{l^2+l+2}{2}\frac{p_b^2}{L_o^2}\bigg)^{-2},
        \end{aligned}\\
        &\begin{aligned}
            \mathfrak{C} &= \bigg[2\Omega_b + 4\bigg((l+2)(l-1)\frac{p_b^2}{L_o^2}\bigg)^{-1}\bigg(\Omega_b\omega_n^2p_c^2 - \Omega_c\frac{l^2+l+2}{2}\frac{p_b^2}{L_o^2}\bigg)\bigg]\bigg(\omega_n^2p_c^2 - l(l+1)\frac{p_b^2}{L_o^2}\bigg)\\
            &\times\bigg(\frac{l^2+l+2}{2}\frac{p_b^2}{L_o^2}\bigg)^{-1} - 8\Omega_b^2\frac{(l-2)!}{(l+2)!}\frac{L_o^4}{p_b^4}\bigg[\bigg(\omega_n^2p_c^2 - l(l+1)\frac{p_b^2}{L_o^2}\bigg)^2 + (l+2)(l-1)\frac{p_b^2}{L_o^2}\omega_n^2p_c^2\bigg]\\
            &\times\bigg(\Omega_b\omega_n^2p_c^2 - \Omega_c \frac{l^2+l+2}{2}\frac{p_b^2}{L_o^2}\bigg)\bigg(\frac{l^2+l+2}{2}\frac{p_b^2}{L_o^2}\bigg)^{-2}.
        \end{aligned}
    \end{aligned}
\end{equation}
We note that these expressions are polynomials in the $\Omega$ variables and the triad variables, except for divisions by strictly positive functions of $p_b$. Since $l\geq 2$ in our discussion, it is straightforward to see that $\mathfrak{A}$ and $\mathfrak{B}$ are positive definite.

\end{document}